\documentclass[aps,pra,reprint,longbibliography,superscriptaddress]{revtex4-1}
\usepackage{graphicx}  
\usepackage{dcolumn}   
\usepackage{bm}        
\usepackage{amssymb}   
\usepackage{verbatim}  
\usepackage{epstopdf}
\usepackage{color}
\usepackage{soul}
\usepackage{amsmath}
\usepackage{subfigure}

\newcommand{\beq}{\begin{equation}}
\newcommand{\eeq}{\end{equation}}
\newcommand{\bea}{\begin{eqnarray}}
\newcommand{\eea}{\end{eqnarray}}

\newcommand{\eq}[1]{{(\ref{#1})}}
\newcommand{\commentout}[1]{{}}

\newcommand{\half}{{\hbox{$\frac{1}{2}$}}}

\newcommand{\bE}{{\bf E}}
\newcommand{\br}{{\bf r}}
\newcommand{\bX}{{\bf X}}
\newcommand{\ave}[1]{{\left<#1\right>}}

\newcommand{\dip}{{\cal D}}
\newcommand{\G}{{\sf G}}
\newcommand{\bd}{{\bf d}}
\newcommand{\bP}{{\bf P}}
\newcommand{\bD}{{\bf D}}

\begin{document}

\title{Exact electrodynamics versus standard optics for a slab of cold dense gas}

\author{Juha Javanainen}
\affiliation{Department of Physics, University of Connecticut, Storrs, Connecticut 06269-3046}
\author{Janne Ruostekoski}
\affiliation{Mathematical Sciences, University of Southampton,
Southampton, SO17 1BJ, UK}
\author{Yi Li}
\affiliation{Department of Physics, University of Connecticut, Storrs, Connecticut 06269-3046}
\altaffiliation[optional text]
    {affiliation information}
\author{Sung-Mi Yoo}
\affiliation{Department of Liberal Arts, Hongik University, 94 Wausan-ro, Mapo-gu, Seoul 04066, South Korea}
\affiliation{School of Computational Sciences, Korea Institute for Advanced Study, 85 Hoegiro, Dongdaemun-gu, Seoul 02455, South Korea}
\affiliation{Department of Physics, University of Connecticut, Storrs, Connecticut 06269-3046}
\begin{abstract}
We study light propagation through a slab of cold gas using both the standard electrodynamics of polarizable media, and massive atom-by-atom simulations of the electrodynamics. The main finding is that the predictions from the two methods may differ qualitatively when the density of the atomic sample $\rho$ and the wavenumber of resonant light $k$ satisfy $\rho k^{-3}\gtrsim 1$. The reason is that the standard electrodynamics is a mean-field theory, whereas for sufficiently strong light-mediated dipole-dipole interactions the atomic sample becomes correlated. The deviations from mean-field theory appear to scale with the parameter $\rho k^{-3}$, and we demonstrate noticeable effects already at $\rho k^{-3} \simeq 10^{-2}$. In dilute gases and in gases with an added inhomogeneous broadening the simulations show shifts of the resonance lines in qualitative agreement with the predicted Lorentz-Lorenz shift and ``cooperative Lamb shift'', but the quantitative agreement is unsatisfactory. Our interpretation is that the microscopic basis for the local-field corrections in electrodynamics is not fully understood.
\end{abstract}
\maketitle
\section{Introduction}
With laser cooling and trapping, and evaporative cooling, it is now experimentally possible to prepare what is arguably the most elementary medium for light propagation, atoms effectively at standstill. More specifically, a cold enough gas presents what according to the long-standing terminology of laser spectroscopists is a homogeneously broadened medium; the atoms move only a small fraction of the wavelength over the time it takes their internal state to relax to steady state. Each atom in the sample is subject not only to the driving light but also to the light sent by all other atoms. On the microscopic level this makes the problem of light propagation in the medium a major challenge. This  subject is obviously very old, but cold atomic samples afford an opportunity for experiments in unprecedented regimes and in unprecedented detail. Correspondingly, new experiments are emerging rapidly~\cite{Meir2013,Pellegrino2014a,wilkowski2,Ye2016,Guerin_subr16,vdStraten16,Jenkins_thermshift,Jennewein_trans,Roof16,Araujo16}.

On the theoretical side, there is the old idea that one could solve the problem of light propagation in a medium on an atom-by-atom basis directly numerically~\cite{Javanainen1999a}. The growing throughput of computers available to researchers is  making such a plan practical. These methods, whether called classical-electrodynamics simulations or coupled-dipole simulations, are now a routine theoretical tool~\cite{CHO12,Jenkins_2d_lat,Bienaime2013,Javanainen2014a,Pellegrino2014a,JavanainenMFT,Sutherland16forward,Ye2016,Bettles_lattice,antoine_polaritonic,YOO16,Jenkins_thermshift,ZHU16,Sutherland1D,Bettles1D,Jennewein_trans}. Closely related numerical techniques based on the analysis of the eigenstates of the coupled system of the light and the atoms~\cite{Rusek96,JenkinsLongPRB,Jenkins_2d_lat,Castin13,Balik2013,Skipetrov14,Facchinetti}
or density matrices and quantum trajectories~\cite{BerhaneKennedyfer,Clemens2003a,Olmos16}
are also widely used today. Other ideas drawn from the theory of radiative transfer~\cite{Chandraskhar1960,Ishimaru1978} and multiple scattering~\cite{Lagendijk,Rossum}, amended with numerics, also have potential to make inroads into the questions about light propagation in atomic media~\cite{Guerin16b}.

The present work started with our chance observation in numerical light propagation simulations that the density-dependent Lorentz-Lorenz (LL) shift of the atomic resonance~\cite{LOR11}, a quintessential local-field correction, is absent in cold, dense atomic samples~\cite{Javanainen2014a}. Delving into the problem deeper, we discovered that the standard electrodynamics of polarizable media (EDPM)~\cite{Jackson,BOR99}, and the resulting standard optics, may  fail qualitatively in cold, dense atomic samples~\cite{JavanainenMFT}. However, adding inhomogeneous broadening that mimics the Doppler shifts of thermal atoms restored the behavior of standard EDPM.
Along the way we made a number of additional qualitative and quantitative observations. Among others, in our analysis it emerged that the so-called ``cooperative Lamb shift'' in a slab of atomic matter~\cite{Friedberg1973} can be explained in standard electrodynamics as an etalon effect due to the reflections of light from the faces of the slab of matter. In retrospect the issues with the EDPM are not much of a surprise, as EDPM is an effective-medium mean-field theory (MFT) and is bound to fail when the light-mediated dipole-dipole interactions make the atomic sample strongly correlated. Nonetheless, they beg for questions about the meaning of cooperative light-atom interactions~\cite{JavanainenMFT,Guerin16b} and the limits of the predictive power of EDPM.

In this paper we add technical details and new results and discussions that illuminate, support and expand on the observations in~Refs.~\cite{Javanainen2014a,JavanainenMFT}. We begin in  Sec.~\ref{BACKGROUND} by reviewing the theoretical basis of our classical-electrodynamics simulations. In Sec.~\ref{SEXP} we orient the reader to concepts such as coherent and incoherent scattering and cooperative line shifts and linewidths  by presenting simple analytically solvable examples. The core of the present paper, however, is about the comparison of the EDPM solutions and numerical simulations of the response of a gas of atoms confined to a slab to light at normal incidence. What exactly is involved here is explained in Sec.~\ref{THESLAB}. The remaining sections~\ref{RESULTS} and~\ref{REMARKS} present and discuss the results.

\section{Background}
\label{BACKGROUND}
The purpose of the present Sec.~\ref{BACKGROUND} is threefold. First we briefly summarize our fully quantum mechanical approach to light-matter interactions as in Refs.~\cite{Ruostekoski1997a,Ruostekoski1997b}, especially as it comes to a hierarchy of equations of motion for the correlation functions that involve polarization of the atoms and densities of the atoms in different points in space. Second, following Ref.~\cite{Javanainen1999a}, we explain how and in what sense we may solve the hierarchy for the correlation functions numerically using classical-electrodynamics simulations. Third, both in our numerical computations and in the discussions of this paper we almost exclusive use certain natural units for microscopic theory of light propagation in dipolar samples. We conclude by introducing these units.

\subsection{Quantum theory of light propagation in dipolar medium}\label{QUANTUMTHEORY}

Our approach~\cite{Ruostekoski1997a} begins with the boson field operators for ground-state and excited-state atoms $\psi_g(\br)$ and $\psi_e(\br)$. The labels $g$ and $e$ implicitly include the Zeeman state labels of the angular-momentum degenerate energy levels. We adopt a summation convention whereby repeated indices $g$ and $e$ in a product are summed over. Since we always deal with pairs of atom field operators, we believe that our scheme is also valid for fermionic atoms.

The atoms are coupled to the quantized electrodynamic field via the dipole interaction. We deviate from the dominant practice  in that we adopt the Power-Zienau-Woolley viewpoint~\cite{PowerZienauPTRS1959,PowerBook,CohenT}, whereby the primary quantized variable having to do with the electromagnetic field is the electric displacement $\hat\bD$ not the electric field $\hat\bE$. The result is a quantum field theory that in appearance closely resembles the usual EDPM.  However, here we deliberate phrase our arguments in terms of the electric field.

To begin with, we have the positive frequency part of the polarization operator for the atoms
\beq
\hat\bP(\br) = \bd_{ge}\psi^\dagger_g(\br)\psi_e(\br),
\eeq
where $\bd_{ge}$ are the dipole moment matrix elements. If there is a difference between positive- and negative-frequency parts of the quantity in question, we write down the positive-frequency part without further comment. Correspondingly, when we consider analogous classical quantities, we alway assume a dominant frequency of the driving light $\omega$ in the problem, so that a classical counterpart of a positive-frequency part of a quantity may be written as, say, $\langle{\hat\bP(\br,t)}\rangle = e^{-i\omega t} \bP(\br,t)$, where $\bP(\br,t)$ now assumedly varies little over the time scale $\omega^{-1}$.  We then express the classical polarization (and electric field, and dipole moment, and so on) in terms of the slowly varying part $\bP(\br,t)$ without further ado. The physical polarization, a real quantity, is $\half \bP(\br,t)e^{-i\omega t} + {\rm c.c.}$. This convention is, of course, deeply ingrained in optical physics and quantum optics.

In analogy to classical electrodynamics, the electric field operator is related to the polarization operator by
\beq
\hat\bE(\br) = \hat\bE_0(\br) + \int d^3r'\,\G(\br,\br') \hat\bP(\br'),
\label{DISPLACEMENT}
\eeq
where $\hat\bE_0\equiv\hat\bD_0/\epsilon_0$ is the electric field in the absence of matter, and
$\G(\br,\br')$, a $3\times3$ matrix, is the dipole propagator such that $\G(\br,\br') \bd$ is the electric field at $\br$ from an oscillating dipole moment $\bd$ at $\br'$~\cite{Jackson,BOR99}. The dipole propagator should also include a singularity term $-\frac{1}{3}\delta(\br-\br')/\epsilon_0$~\cite{Jackson} in order for \eqref{DISPLACEMENT} to be the integral representation of the correct Maxwell's wave equation, although its presence in the equations of motion for matter is a subtle matter~\cite{Ruostekoski1997b} to which we briefly return later. Integrals involving $\G$ are typically not absolutely convergent either at small or large $|\br-\br'|$. The values of such integrals depend on how they are done. This type of ambiguities are widespread in the theory of the electrodynamics of dipolar media, and are often difficult to resolve.  This is one of the reasons why we think that one should be suspicious of any and all ``physical'' approximations in this field.

One can have the electric field radiated by matter fall back on the matter and change the atomic dipole moments, hence  polarization. The self-field giving rise to radiation reaction and transition linewidths can be handled with the Markov and Born approximations of quantum optics as usual, but otherwise the ensuing operator equations are (most likely) impossible to solve directly. Instead we go to expectation values.

Here we proceed under the limit of low light intensity, only keeping the leading nontrivial contribution in the strength of the incoming field $\hat\bE_0$, and specialize to the case when the angular momenta of the levels are $J_g=0$ and $J_e=1$, c.f.\ Ref.~\cite{Lee16}. Specifically, introduce normally ordered correlation functions for ground state density and correlations between polarization and ground state density as
\bea
\rho_1(\br_1) &=& \ave{\psi^\dagger_g(\br_1)\psi_g(\br_1)}\equiv\rho(\br_1),\nonumber\\
\rho_2(\br_1,\br_2) &=& \ave{\psi_g^\dagger(\br_1)\psi_g^\dagger(\br_2)\psi_g(\br_2)\psi_g(\br_1)} ,\nonumber\\
&&\ldots\,;\\
\bP_1(;\br_1) &=&\langle{\hat\bP(\br_1)}\rangle=\ave{\bd_{ge}\psi^\dagger_g(\br_1)\psi_e(\br_1)}\equiv \bP(\br_1),\nonumber\\
\bP_2(\br_1;\br_2) &=&\langle{\psi_g^\dagger(\br_1)\hat\bP(\br_2)\psi_g(\br_1)}\rangle,\nonumber\\
\bP_3(\br_1,\br_2;\br_3) &=&\langle{\psi_g^\dagger(\br_1)\psi_g^\dagger(\br_2)\hat\bP(\br_3)\psi_g(\br_2)\psi_g(\br_1)}\rangle,\nonumber\\
&&\ldots \,.
\eea
A rigorous quantum mechanical analysis~\cite{Ruostekoski1997a} finds a hierarchy of equation of motions for these expectation values beginning with
\begin{widetext}
\bea
\dot\bP(\br_1) &=& (i\Delta-\gamma)\bP(\br_1)+i\zeta \bE_0(\br_1)\rho(\br_1) +i\zeta\int d^3r_2 \G(\br_1,\br_2) \bP_2(\br_1;\br_2),\label{POLEQ}
\eea
\bea
\dot\bP_2(\br_1;\br_2) &=& (i\Delta-\gamma)\bP_2(\br_1;\br_2)
+i\zeta \bE_0(\br_2)\rho_2(\br_1,\br_2)+ i\zeta \G(\br_2,\br_1)\bP_2(\br_2;\br_1)\nonumber\\&&+i\zeta \int d^3r_3\G(\br_2,\br_3) \bP_3(\br_1,\br_2;\br_3)
\label{POLCOREQ}\,,
\eea
\bea
\dot\bP_3(\br_1,\br_2;\br_3) &=& (i\Delta-\gamma)\bP_3(\br_1,\br_2;\br_3)
+i\zeta \bE_0(\br_3)\rho_3(\br_1,\br_2,\br_3)+ i\zeta \G(\br_3,\br_1)\bP_3(\br_2,\br_3;\br_1)+ i\zeta \G(\br_3,\br_2)\bP_3(\br_1,\br_3;\br_2)\nonumber\\
&&+i\zeta \int d^3r_4\G(\br_3,\br_4) \bP_4(\br_1,\br_2,\br_3;\br_4)
\label{POLPOLCOREQ}\,,
\eea
\end{widetext}
and continuing along these lines all the way up to the order equal to the number of the atoms $N$.
Here the detuning $\Delta=\omega-\omega_0$ is the difference of the frequency of the driving light $\omega$ from the atomic resonance frequency $\omega_0$ and $\gamma$ is the HWHM linewidth of the transition. Further, we have
$\zeta = \dip^2/\hbar$, where $\dip$ is the reduced dipole matrix element related to the linewidth and the wave number of resonant light $k_0=\omega_0/c$ by
\beq
\gamma = \frac{\dip^2k_0^3}{6\pi\hbar\epsilon_0}\,.
\label{GAMMA}
\eeq
In the usual way we assume that the incoming quantum light is in a coherent state, and replace it with the classical electric field $\bE_0$.

Let us factor in Eq.~\eq{POLEQ} the second-order correlation function $\bP_2$ as in
\beq
\bP_2(\br_1;\br_2) = \rho(\br_1)\bP(\br_2).
\eeq
The resulting approximate equation,
\bea
\dot\bP(\br_1) &=& (i\Delta-\gamma)\bP(\br_1)+i\zeta \bE_0(\br_1)\rho(\br_1)\nonumber\\
&&+i\zeta\rho(\br_1)\int d^3r_2 \G(\br_1,\br_2) \bP(\br_2),
\label{MFT}
\eea
tells us that the polarization at $\br_1$, basically the dipole moment of an atom at $\br_1$, evolves under the joint influence of the incoming field and the electric field radiated from the polarization of the atoms, as if the atoms were smeared out continuously in space. This is an effective-medium MFT. Moreover, it is easy to see in explicit examples that this MFT is the same as the standard EDPM for the atoms.

The equation of motion of the second-order correlation function $\bP_2(\br_1;\br_2)$ has a similar structure except for one crucial point, the term on the right-hand side $\propto \bP_2(\br_2;\br_1)$. This is the effect of the dipolar field of the atom at the position $\br_1$ on the atom at $\br_2$. The equation for $\bP_2(\br_2;\br_1)$ has an analogous term, effect of the dipolar field of the atom at the position $\br_2$ on the atom at $\br_1$. This is our first glimpse of recurrent scattering, repeated photon exchange between a group of atoms; in this case, two atoms. The equation for the correlation function $\bP_3(\br_1,\br_2;\br_3)$ similarly exhibits recurrent scattering between atoms at $\br_1$, $\br_2$ and $\br_3$, and so forth.

We also argued in Ref.~\cite{Ruostekoski1997b} essentially as follows: First, assume that the density equals a constant $\rho$, and all position correlations factor as $\rho_n=\rho^n$. Second, state an ansatz for all position-polarization correlations $\bP_k(\br_1,\ldots,\br_{k-1};\br_k)= \rho^{k-1}\bP(\br_k)$. Third, ignore all recurrent-scattering cross terms such as the direct coupling of $\bP_2(\br_2;\br_1)$ to $\bP_2(\br_1;\br_2)$. Then $\bP(\br)$ from Eq.~\eq{MFT} gives the {\em exact\/} solution to the entire hierarchy of the equations of the correlation functions. While this was not claimed in~Ref.~\cite{Ruostekoski1997b}, we believe that the same argument is valid even if the density is not constant as long as all density correlation functions factorize to a product of one-particle densities. Thus, there would be two possible reasons for beyond-MFT effect: Repeated exchange of photons between the atoms (recurrent scattering), and pre-existing correlations in the positions between the atoms.

Another mathematical point we made in Ref.~\cite{Ruostekoski1997b} about the cross terms runs as follows: Since the dipolar kernel $\G(\br_1,\br_2)$ diverges for $\br_1=\br_2$, it follows from (the steady-state version of) Eq.~\eq{POLCOREQ} that $P_2(\br_1;\br_2)$ tends to zero when $\br_1$ and $\br_2$ tend to the same value, and this observation is independent of whether the delta function divergence is present in $\G(\br_1,\br_2)$. Therefore the delta function divergence in $\G$, if any, should have no effect on $\bP(\br_1)$ solved from Eq.~\eq{POLEQ}. By an analogous argument, the delta function divergence has no effect on $\bP_2(\br_1;\br_2)$, and so forth. In short, any delta function divergence in $\G$ should have no effect on the polarization of the gas. This observation is, coincidentally, compatible with the physical notion that it is impossible to overlay two atomic dipoles exactly, so the delta function in $\G$ should never fire. Put  mathematically, the actual position correlations $\rho_k$ for $k\ge2$ should tend to zero when any two of the positions get close. One then surmises from the hierarchy that all correlation functions $\bP_k$ for $k\ge2$ have the same property.

Given the solution to the hierarchy of the equations of the correlation functions, one may obtain the expectation value of the electric field $\bE(\br) = \langle{\hat\bE(\br)}\rangle$ from Eq.~\eq{DISPLACEMENT},
\beq
\bE(\br) = \bE_0(\br) +\int d^3r'\,\G(\br,\br') \bP(\br').
\label{E-EVERYWHERE}
\eeq

Notably absent from the formulation is the motion of the atoms. At the quantum level one may treat the center-of-mass motion of the atoms as a quantized degree of freedom, as one often does in the theory of the mechanical effects of light. We have not done so, however, as the ensuing theory would be cumbersome and opaque. The point to remember here is that the motion of the atoms, forces of light, and effects of photon recoil are all ignored in our present analysis.

\subsection{Classical-electrodynamics solution for light propagation}\label{SIMULATIONS}

Obviously it is in general impossible to solve the hierarchy for polarization correlation functions starting with Eqs.~\eq{POLEQ}-\eq{POLPOLCOREQ}, and so on, directly numerically. The present subsection is mostly a brief summary of Refs.~\cite{Javanainen1999a,Lee16} that present a workaround.  The idea is that, just as one might solve a Fokker-Planck equation (diffusion equation) numerically using stochastic Langevin equations for individual particles~\cite{GardinerBook}, one may solve for the polarization correlations using stochastic classical-electrodynamics simulations.

Consider a gas of $N$ atoms. The simulations start with generation of random positions $\bX_1$, \ldots, $\bX_N$ for the atoms in such a way that the  positions are drawn from the probability distribution that gives the prescribed density correlations $\rho_k$. Incidentally, the normally ordered quantum correlations correspond to the properly defined classical density correlation functions for point-like particles from which the singularities corresponding to counting of the same particle more than  once have been eliminated. For instance, the classical two-particle correlation function would be the stochastic average
\beq
\rho_2(\br_1,\br_2) = \ave{\sum_{i\ne j}\delta(\br_1-{\bf X}_i) \delta(\br_2-{\bf X}_j)},
\eeq
where the restriction $i\ne j$ removes the singularity.

Four cases of atomic probability distribution have been relevant in our work. First, we deal with classical atoms that are assumedly distributed completely independently of one another with the given density $\rho(\br)$. This strictly speaking cannot be true, as, for instance, atoms attract or repel each other at short distances, and the positions of the atoms may also be correlated on the length scale of the thermal de Broglie wavelength. However, we assume that the characteristic distance between the atoms is much larger than this type of correlations lengths. Second, maybe paradoxically, the same independent-atom model applies as the leading approximation also to the Bose-Einstein condensate. Third, atoms confined in optical lattices provide structured arrays where the positions of atoms in the Mott-insulator states can be sampled~\cite{Jenkins_2d_lat,YOO16}. Fourth,
we have also done one-dimensional simulations of a noninteracting one-component Fermi gas at zero temperature~\cite{Javanainen1999a,RUO16}. The joint probability density for the positions of the atoms is then given by the absolute square of the many-body wavefunction, a Slater determinant. The characteristic feature of the Fermi-Dirac statistics is that the atoms tend to avoid each other, and end up with more evenly-spaced positions than classical atoms.  Atoms with this position distribution may be sampled using the Metropolis algorithm.

Given the positions of the atoms, corresponding to the hierarchy for the correlation functions we next have the equations of motion for the dipole moments of the atoms. Denoting the positions of the dipoles explicitly, we have
\bea
&&\dot\bd(\bX_i)\!=\! (i\Delta\!-\!\gamma)\bd(\bX_i)\!+\!i\zeta\bE_0(\bX_i)\!+\!i\zeta\! \sum_{j\ne i}\! \G(\bX_i,\!\bX_j)\bd(\bX_j).\nonumber\\
\label{DIPOLEEQS-T}
\eea
The dipole at $\bX_i$ is driven by the external field $\bE_0$, and by the dipolar fields from all other atoms. This may be seen even more graphically from the steady-state version of Eqs.~\eq{DIPOLEEQS-T},
\beq
\bd(\bX_i)= \alpha\left(\bE_0(\bX_i) + \sum_{j\ne i} \G(\bX_i,\bX_j)\bd(\bX_j)\right)\,,
\label{DIPOLEEQS-S}
\eeq
where
\beq
\alpha = - \frac{\zeta}{\Delta+i\gamma} = -\frac{\dip^2}{\hbar}\frac{1}{\Delta+i\gamma}
\label{POLARIZABILITY}
\eeq
is the polarizability of an atom; in fact, the well-known polarizability of the proverbial two-level atom at low light intensity. In the present case the steady-state dipole moment aligns with the net field at the position of the atom. This isotropy is because of our underlying assumption of the $J_g=0\rightarrow J_e=1$ transition. However, for other types of transitions~\cite{JEN_longpaper} the appropriate polarization tensor $\alpha_{ij}$ could be defined such that the relation between the vector components of the the dipole moment and the electric field reads $d_i = \sum_{j}\alpha_{ij}E_j$.

In the present paper we deal solely with the steady-state version of the theory, Eqs.~\eq{DIPOLEEQS-S}.  These are a closed inhomogeneous set of linear equations for the dipole moments, or, thinking about it in another way, for the electric fields at the positions of the dipoles $\bE(\bX_i)$:
\beq
\bE(\bX_i) = \bE_0(\bX_i) +\alpha \sum_{j\ne i} \G(\bX_i,\bX_j)\bE(\bX_j)\,.
\label{FIELDSONDIPOLES}
\eeq
Given a sample of the positions of the atoms, we solve these equations numerically for $\bE(\bX_i)$. Regarding the analogy to solving diffusion equations using particle simulations, this solution would be the counterpart of a stochastic trajectory obtained from the Langevin equation.

In the end we are interested in the total electric (and possibly also magnetic) field everywhere in space. It is formally given by
\beq
\bE(\br) = \bE_0(\br) + \alpha \sum_{j} \G(\br,\bX_j)\bE(\bX_j)\,
\label{FIELDFROMDIPOLES}
\eeq
everywhere except at the exact positions of the atoms, where we have a divergence in the dipolar kernel. In an indirect way, even this divergence has been taken into account: The dipolar field acting back on the atom that sends the field is formally infinite, but the action of this self-field is already included in the damping rate $\gamma$, and the associated level shift (Lamb shift) is incorporated into the energies of the levels. We also caution that the electric field as in Eq.~\eq{FIELDFROMDIPOLES} is not necessarily the most practical quantity to calculate. We return to this point below in Sec.~\ref{NUMSIM}.

The remaining step in the simulations is to repeat the process for a large number of samples of the atomic positions, and average the results. In the limit of an infinite number of samples, the stochastic average of, say, the electric field converges to the corresponding quantum mechanical average that would be obtained by solving the entire quantum hierarchy for the correlation functions.

Several analyses assuming that there is at most one photon present at any time~ \cite{PRA00,SVI10,Balik2013,Bienaime2013} also in effect show that in the limit of low light intensity quantum theory of light-matter interactions reduces to classical electrodynamics. There has been an argument along these lines that found some deviations from the standard EDPM~\cite{SVI16}; specifically, a result that was traditionally thought to apply for the displacement was derived for the electric field. We emphasize, though, that by strictly following the Power-Wolley-Zienau procedure~\cite{PowerZienauPTRS1959,PowerBook,CohenT}, with the inclusion of the polarization self-energy, we got results that were in a complete agreement with the structure of the usual EDPM~\cite{Ruostekoski1997a}; compare Eq.~(13) of Ref.~\cite{Ruostekoski1997a} and Eq.~(14) of Ref.~\cite{SVI16} with Eq.~(31) of Ref.~\cite{SVI16}. There are also caveats to the quantum-classical agreement, cases such as 1D nanofibers or photonic crystals~\cite{Thompson07062013,kimblemanybody} in which the atom-field coupling can be so strong that one photon may saturate an atom. We will not analyze such situations any further in the present paper.

From this point on we leave quantum mechanics behind, and pretend that the electrodynamics of the dipolar medium is, in fact, entirely classical. For instance, for any given atomic sample we find the electric field which we may average over many samples to get the averaged field. Likewise, given the electric field, we square it, average, and obtain the average of the square of the electric field, which is an intensity-like quantity. This is not a trivial point in quantum mechanics: If we were to compute the quantum average of $\hat\bE^\dagger(\br)\cdot\hat\bE(\br)$ exactly from quantum mechanics, we would in principle have to do something like develop a hierarchy of equations for {\em some other\/} correlation functions than those we have dealt with so far, and, if possible, develop a corresponding simulation. The difference between the quantum mechanical average of the square of the quantum field and the classical average of the square of the classical field is in the quantum fluctuations.

Nonetheless, we do not expect significant quantum fluctuations in the kind of situations we consider here. In fact, for a model atom such as ours, in the limit of low light intensity, there are no quantum fluctuations in the scattered light. We may have an issue for instance if the light intensity is increased, or if more than one electronic Zeeman ground level is involved~\cite{Lee16,HAM10}, although observing such quantum fluctuations typically require
delicate experimental setups.

\subsection{Units and conventions}

All numerical computations described here were done in units such that the numerical values
\beq
k = c = \hbar = \frac{1}{4\pi\epsilon_0} = 1
\eeq
apply. Here $k=\omega/c$ is the wave number of the driving light. For parameters typical in laser spectroscopy the difference between the wave number of the driving light and of resonant light $k_0=\omega_0/c$ is negligible, and we henceforth ignore it. The unit of length $k^{-1}$ is related to the wavelength of the driving light by $k^{-1}=\lambdabar=\lambda/2\pi$, and the unit of quantities such as area and density follow accordingly. Below all discussions are in these units, unless explicitly stated otherwise.

For a dipole $\bd$ at ${\bf r}_0$, the electric and magnetic fields of dipole radiation at position $\bf r$ are
\beq
{\bf E}({\bf r}) =\G({\bf r},{\bf r}_0) {\bf d},\quad
{\bf B}({\bf r}) = {\sf H}({\bf r},{\bf r}_0){\bf d},
\eeq
where $\G$ is again the dipolar field propagator, and $\sf H$ gives the magnetic field from a dipole. Expressed in Cartesian coordinates, these are matrices with the  components
\bea
&&\G_{ij}({\bf r},{\bf r}_0) =\nonumber\\ &&\hat{\bf e}_i \cdot\left\{
(\hat{\bf n}\times \hat{\bf e}_j)\times\hat{\bf n} + [3\hat{\bf n}(\hat{\bf n}\cdot\hat{\bf e}_j)-\hat{\bf e}_j]\left( \frac{1}{r^2}-\frac{i}{r}\right)\right\}\frac{e^{ir}}{r}\,,\nonumber\\
\label{DIMLESSG}\\
&&{\sf H}_{ij}({\bf r,\br_0}) =\hat{\bf e}_i \cdot\hat{\bf n}\times\hat{\bf e}_j\left(1-\frac{1}{ir}\right)\,\frac{e^{ir}}{r}\,.\label{DIMLESSH}
\eea
Here $r$ and $\hat{\bf n}$ are the distance from the source point to the field point and the unit vector directed from the source point to the field point, and $\hat{\bf e}_i$ are the cartesian unit vectors.
We have dropped the contact term in \eqref{DIMLESSG} as we here observe the light outside the sample, and in the interactions between the atomic dipoles it is inconsequential~\cite{Ruostekoski1997b}.
The relation between the positive frequency parts of the electric and magnetic fields reads
\beq
{\bf B}({\bf r}) =-i\,\nabla\times{\bf E}({\bf r}),
\eeq
and the energy density and Poynting vector at the given field position are
\bea
\hbox{\sc e} &=& \frac{1}{16\pi}\, ({\bf E}\cdot{\bf E}^*+{\bf B}\cdot{\bf B}^*),\\
{\bf S} &=& \frac{1}{8\pi}\, \Re[{\bf E}\times{\bf B}^*]\,.
\eea

Another convention here is that, by default, we express the detuning in units of the linewidth of the transition,
\beq
\Delta = \delta\gamma,
\eeq
with the dimensionless detuning $\delta$.
Be virtue of Eq.~\eq{GAMMA} and our conventions, the relation between dipole moment matrix element and linewidth reads
\beq
\dip =\sqrt\frac{3\gamma}{2},
\eeq
and the polarizability~\eq{POLARIZABILITY} is
\beq
\alpha = - \frac{3}{2} \frac{1}{\delta+i}\,.
\label{DIMLESSALPHA}
\eeq

Assuming a single atom at the origin and an incoming field with the vector amplitude $\bE_0$, one may straightforwardly obtain the dipole moment, the electric and magnetic fields, the Poynting vector, and finally the radiated power. The result, a useful reference, is 
\beq
P = \frac{|\alpha|^2|\bE_0|^2}{3}=\frac{3|\bE_0|^2}{4(1+\delta^2)}.
\eeq
The intensity of the incoming plane wave is
\beq
I_0=\frac{1}{8\pi}|\bE_0|^2.
\eeq
 Writing the radiated power in terms of the intensity and the scattering cross section $\sigma$ as $P=\sigma I_0$, we find
\beq
\sigma(\delta)=\frac{8\pi}{3}|\alpha|^2=\frac{6\pi}{1+\delta^2}\,.
\label{INDATSC}
\eeq
The on-resonance light scattering cross section therefore is $\sigma(0)=6\pi$.

\section{Simple examples}\label{SEXP}
In this section we discuss independent-atom response and cooperative response to light in simple analytically solvable cases. Among other things we demonstrate the difference between coherent and incoherent scattering, and show how even the seemingly starkly contrasting concepts of cooperativity from Dicke states and radiation from independent atoms may be difficult to tell apart. Our examples serve as a general reference for more comprehensive numerical-simulation studies that follow.  We also introduce several pieces of physics that are absent from the present simplest models, but might well figure in real experiments and complicate the interpretation of simulation results.

\subsection{Radiation from a Gaussian clouds of atoms}

\subsubsection{Continuous medium}

For the problem of $N$-atom gases, let us start with a hypothetical model with a continuous spatial distribution of atoms. In terms of macroscopic electromagnetism, there is a monochromatic polarization of the sample
${\bf P}({\bf r}) = \rho({\bf r}) \,{\bf d}({\bf r})$,
where $\rho({\bf r})$ is the density of the sample  and $\bf d({\bf r})$ is the electric dipole of an atom at the position $\bf r$. Taking the atoms  to reside around the origin of the coordinates, in the far field at the distance $r\gg1$ and with $r$ much larger than the size of the sample, the terms  $\propto 1/r^2$ and $\propto 1/r^3$ in Eq.~\eq{DIMLESSG} are negligible and the  field  radiated (``scattered'') by this polarization is
\bea
{\bf E}_S({\bf r})&\simeq&\frac{e^{ir}}{r} \int d^3r'\,e^{-i  \hat{\bf r}\cdot {\bf r}'}\rho({\bf r}')
\, [\hat{\bf r}\times{\bf d}({\bf r}')]\times\hat{\bf r}\,;
\label{SCATFIELD}
\eea
$\hat{\bf r}={\bf r}/r$ is the unit vector that points from the source at $\simeq 0$ toward the field point at the distance~$r$.

For easy analysis, we model the density with a Gaussian,
\beq
\rho({\bf r}) = \frac{3\sqrt3N}{2\sqrt2\pi^{3/2} R^3}\, e^{-\frac{3r^2}{2R^2}}\,,
\label{CONT_DEN}
\eeq
where $N$ is the atom number and $R$ is the size scale of the sample. The parametrization is chosen in such a way that the rms value of $|\br|$ equals $R$. In the limit $R\gg1$ there will be a narrow cone of radiation around the direction of the incoming beam; let us denote the angle from the incident beam by $\theta$.

For a tangible example we take a $\sigma_+$ circularly polarized plane wave propagating in the $z$ direction, writing
\beq
\bE_0(\br) = E_0\, e^{iz}\,\hat{\bf e}_+;\quad \hat{\bf e}_+=-\frac{1}{\sqrt{2}}(\hat{\bf e}_x + i \hat{\bf e}_y)\,.
\label{incomingE}
\eeq
The assumption is that the incoming light dominates even inside the sample, i.e., that each atom responds to the incoming light only. Accordingly, we write the dipole moment of an atom at $\br$ as
\bea
{\bf d}({\bf r})=\alpha\bE_0(\br)=\alpha E_0\, e^{iz}\,\hat{\bf e}_+.
\eea
The radiated field from Eq.~\eq{SCATFIELD} is then
\bea
{\bf E}_S({\bf r})&=&\frac{\alpha NE_0e^{ir-\frac{1}{3}R^2(1-\cos \theta)}}{r}[ (\hat { \bf n} \times \hat{{\bf e}}_+)  \times \hat {\bf n}].
\eea
In the far field the light locally makes a plane wave, the Poynting vector points radially outwards and has the magnitude
\bea
S_S({\bf r})&=&\frac{|{\bf E}_S|^2}{8\pi}=\frac{\left|\alpha\right|^2N^2 E_0^2 e^{-\frac{2}{3}R^2(1-\cos \theta)} (1+\cos^2 \theta)}{16 \pi r^2},\nonumber\\
\eea
and the total power in the radiation is readily obtained as
\bea
P_S&=&\int d^2\Omega\, r^2\, S_S({\bf r})\,\nonumber\\
&=&\!\!\frac{3\left|\alpha\right|^2\!\!N^2\!E_0^2\left[\!(4R^4\!-\!6R^2\!+\!9)\!-\!e^{-\frac{4R^2}{3}}(4R^4\!+\!6R^2\!+\!9)\!\right]}{32R^6}.\nonumber\\
\eea
The intensity of the scattered light scales with the square of the atom number, $N^2$. This is similar to an important characteristic of superradiance. However, it is early for conclusions yet. Instead, we will next inspect a more realistic problem with discrete atoms.

\subsubsection{Independent discrete radiators}
Take a collection of $N$ identical dipoles sitting at the positions ${\bf r}_i$ in the incoming field, and assume that each of these dipoles radiates a field ${\bf E}_i({\bf r})$ {independently}. In other words, we again assume that only the incoming field $\bE_0(\br)$ drives each dipole. In terms of scattering theory, one might say that a photon scatters from an atom at most once, so this model is occasionally called single-scattering approximation.

The total dipolar field at the point ${\bf r}$ is
\beq
{\bf E}_S({\bf r}) = \sum_i {\bf E}_i({\bf r})\,,
\eeq
with
\beq
{\bf E}_i({\bf r}) = \alpha\,\G(\br,\br_i)\bE_0(\br_i)\,.
\eeq
In the far field the radial component of the Poynting vector is
\bea
S_S(\br) &=& \frac{1}{8\pi} \bE_S(\br)\cdot\bE_S^*(\br )\nonumber\\
&=& \frac{1}{8\pi} \sum_{i,j} \bE_i(\br)\cdot\bE_j ^*(\br)\,.
\eea

We take the position of each atom to be a random variable independent of the positions of the other atoms, governed by the probability density function $f(\br)$. Then the average outward energy flux (average over many samples of the gas) is determined from
\bea
8 \pi\bar{S}_S &=&\left\langle
 \sum_{i\ne j} \bE_i(\br)\cdot\bE_j^*(\br)+\sum_i \bE_i(\br)\cdot\bE_i^*(\br)
\right\rangle\nonumber\\
&=&
 \sum_{i\ne j} \left\langle\bE_i(\br)\right\rangle
\cdot\left\langle\bE_j^*(\br)\right\rangle +
\sum_{i} \left\langle\bE_i(\br)\cdot\bE^*_i(\br)\right\rangle\nonumber\\
&=&N(N-1)|\left\langle \bE_i(\br)\right\rangle|^2 + N \left\langle\bE_i(\br)\cdot\bE^*_i(\br)\right\rangle\!.
\eea
The first term represents coherent scattering, as if the atom was spread out to a continuous dielectric material with the spatial shape specified by $f(\br)$. It arises from adding the fields of different radiators, and is essentially proportional to $N^2$. The second term $\propto N$ is for incoherent scattering, the sum of the intensities radiated by the individual atoms. It is present because the gas is not a continuous (nonfluctuating) dielectric medium, but consists of discrete scatterers.

Incidentally, while the above argument might not be as widely known as it deserves to be, the basic message is far from novel. If air were a continuous dielectric medium, it would not scatter sunlight sideways and the sky would be black. The blue sky comes from incoherent scattering that results because air consists of discrete molecules. This observation goes back to (at least) Lord Rayleigh~\cite{Rayleigh1899}.

For a comparison, we apply the same incoming light as in Eq.~\eq{incomingE}, and the position distribution for each atom is taken to be the same Gaussian,
\beq
f(\br) = \frac{3\sqrt3}{2\sqrt2\,\pi^{3/2}R^3}\,e^{-\scriptstyle\frac{3r^2}{2R^2}}\,.
\eeq
Given the usual polarizability $\alpha$, the far field (the $1/r$ part of dipole radiation) averaged over the positions of an atom, the absolute square of the former, and the  absolute square  of the field averaged over the positions give
\bea
\langle \bE_i(\br)\rangle &=& \alpha E_0 e^{-\hbox{$\frac{1}{3}$}R^2(1-\cos\theta)}[(\hat{\bf r}\!\times\!\hat{\bf e}_+)\!\times\!\hat{\bf r}]\,\frac{e^{ir}}{r},\\
|\langle \bE_i(\br)\rangle|^2 &=& \frac{|\alpha |^2| E_0|^2[3+\cos (2 \theta )] e^{-\frac{2}{3} R^2 [1-\cos \theta
   ]}}{4 r^2},\\
   \langle |\bE_i(\br)|^2\rangle &=& \frac{|\alpha |^2| E_0|^2[3+\cos (2 \theta )]}{4 r^2}\,,
\eea
and the total radiated power becomes
\bea
&&P_S=\frac{|\alpha|^2 |E_0|^2}{3}\,\times\nonumber\\
&&
\left(\!\!N(N\!-\!1)\frac{9\left[\!(4R^4\!\!-\!\!6R^2\!\!+\!\!9)\!-\!e^{-\frac{4R^2}{3}}(4R^4\!\!+\!6R^2\!\!+\!\!9)\!\right]}{32R^6}\!+\! N\!\right).\nonumber\\
\label{PN}
\eea

\begin{figure}[h!]
\begin{center}
\includegraphics[width=0.5\columnwidth]{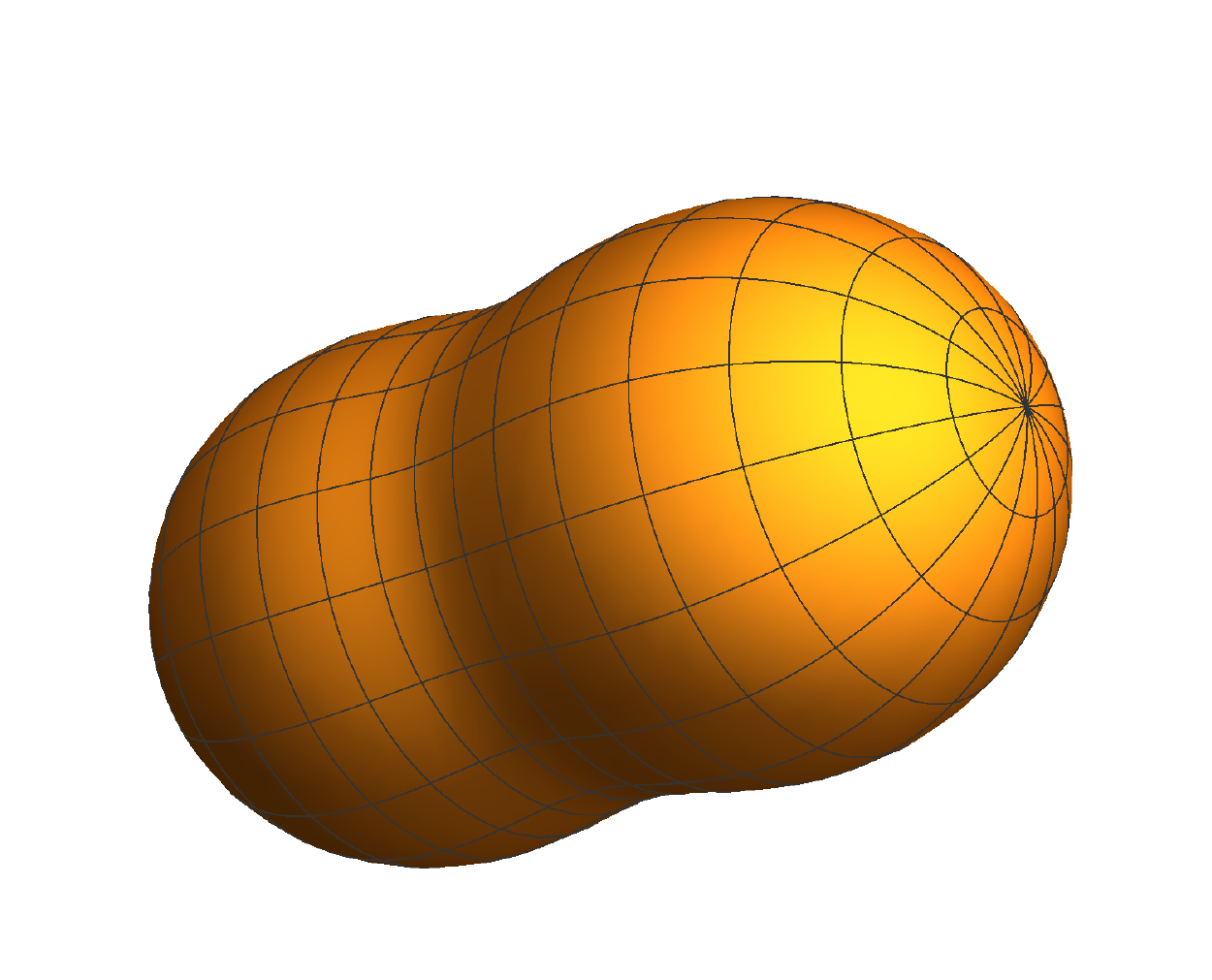}
\hspace{-0.05\columnwidth}\includegraphics[width=0.5\columnwidth]{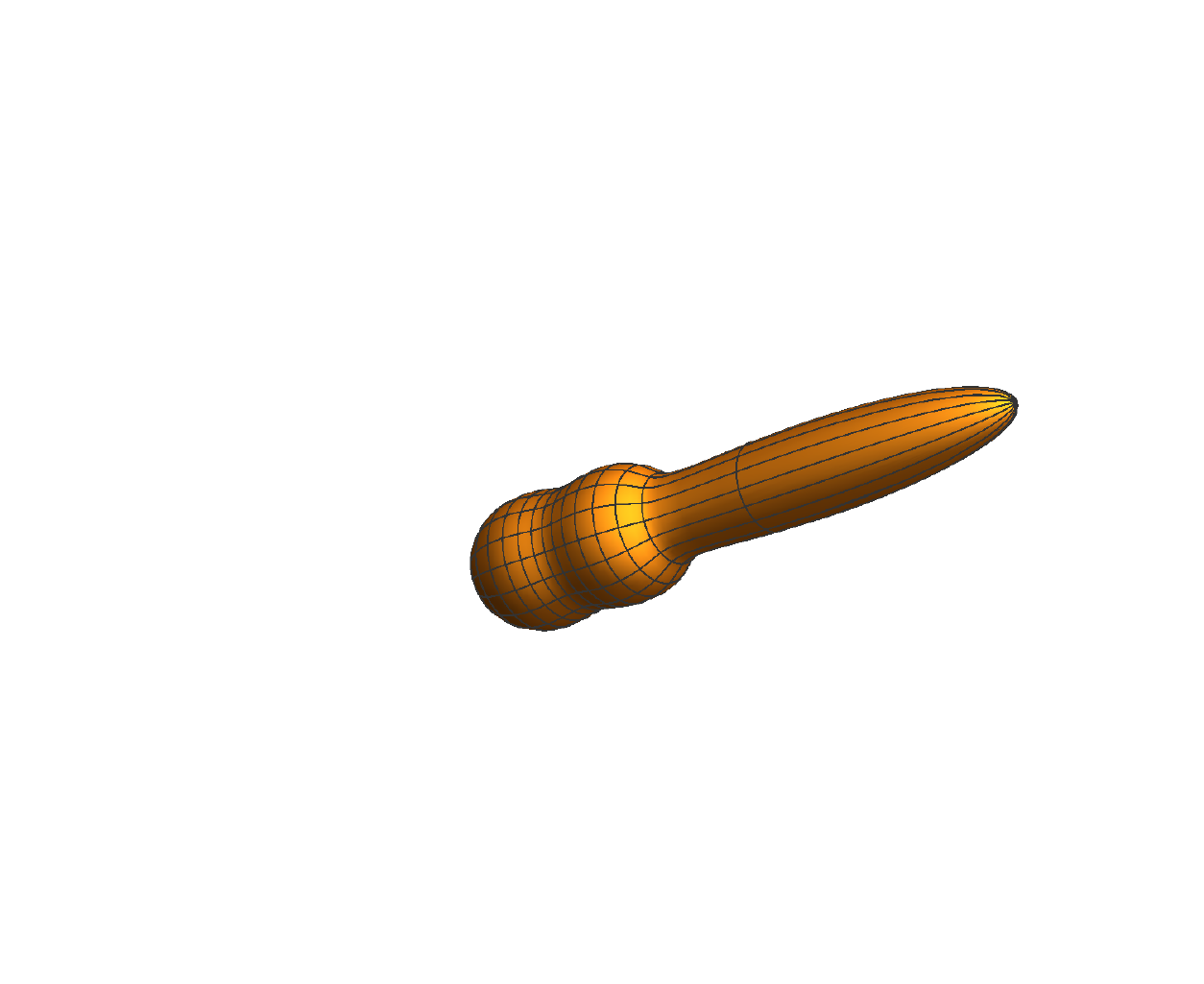}
\end{center}
\caption{Radiation patterns for a Gaussian cloud with $N=4$,  $R=0.1$ (left), and $R=10$ (right) in the single-scattering approximation. The propagation direction of the driving light $z$ is along the long axis of the radiation patterns. The scale is arbitrary but the same for both figures, so the total power is obviously much larger for the cloud with the smaller radius 0.1.}
\label{ANGULAR}
\end{figure}

The part $\propto N(N-1)$ in Eq.~\eq{PN} is the same as it would be for the continuous atom density in Eq.~\eq{CONT_DEN}, except for the factor of $N(N-1)$ instead of $N^2$. In the forward direction $\theta=0$, the intensities of  coherently and incoherently scattered components of light add up to exactly $N^2$ times the intensity from a single atom. The bigger is the sample, the narrower is the cone in the direction $\theta\simeq0$ for coherent scattering.  This aspect is demonstrated in Fig.~\ref{ANGULAR}. The sample acts as an antenna that directs the radiation in the forward direction. The total power of scattered light decreases with increasing size of the cloud.

Conversely, in the limit $R\rightarrow0$ the intensity in all directions is enhanced by the factor $N^2$, and of course so is the total power. Given the well-known Dicke cooperative regime, a reader might erroneously interpret such an enhancement as a cooperative phenomenon. It cannot be, since in this example we have simply added the fields from independent radiators.


\subsection{Radiation from two atoms}
\label{TWOATOMRADIATION}
For two atoms (or ions, as things might be) the radiation field can be solved explicitly, and there have even been experiments already a while ago~\cite{Eichmann1993,DeVoe1996}. We discuss as an example the special case when a plane wave polarized in the $x$ direction and propagating in the $z$ direction strikes two atoms sitting on the $x$ axis separated by the distance $\ell$. Specifically, we have the incoming field and the two positions for the dipoles
\beq
E_0({\bf r}) = E_0\, \hat{\bf e}_x\,  e^{i z},\quad {\bf r}_{\pm} = \pm\half\,\ell\,\hat{\bf e}_x\,.
\eeq

In this case the fields at the positions of the dipoles as solved from Eqs.~\eq{FIELDSONDIPOLES} are
\beq
{\bf E}({\bf r}_\pm) = \frac{\ell ^3}{\ell ^3+2 i \alpha  \ell e^{i \ell }-2 \alpha  e^{i \ell }}\,E_0\,\hat{\bf e}_x\,.
\label{FOD}
\eeq
Since the dipolar field diverges with decreasing distance from the dipole, one might expect that the fields at the positions of the dipoles should diverge when the dipoles approach one another. However, the exact opposite holds true: For a fixed detuning and hence fixed polarizability $\alpha$, ${\bf E}({\bf r}_\pm)$ actually tend to zero as $\ell^3$ when the distance $\ell$ between the dipoles tends to zero. Maybe counterintuitively, when the detuning is kept constant and the atoms approach each other, they decouple from the light altogether~\cite{JEN_longpaper}. That is why we are not overly concerned about some atoms being close to one another in the steady-state numerical simulations.

Fixed detuning, however, may not be the most useful way of viewing the result. Instead, we insert the explicit expression of the polarization. In this subsection~\ref{TWOATOMRADIATION} only, we find it expedient {\em not\/} to scale the detuning to the linewidth $\gamma$, and write
\bea
{\bf E}({\bf r}_\pm) &=& \left\{
1 + \frac{3 e^{i\ell}(1-i\ell)/\ell^3}{\Delta(\ell)-i\gamma(\ell)}
\right\}E_0\,\hat{\bf e}_x;\\
\Delta(\ell) &=& \Delta -3  \left[\frac{\cos (\ell )}{\ell ^3}+\frac{\sin (\ell )}{\ell
   ^2}\right] \gamma,\nonumber\\
\gamma(\ell) &=&\left[1+\frac{3 \sin (\ell )}{\ell ^3}-\frac{3 \cos (\ell )}{\ell ^2}\right]\gamma \,.
\label{GAMMAL}
\eea
This shows our first instance of cooperative shift and broadening of the resonance of the atoms as a result of the radiation from one dipole falling on the other. The sines and cosines originate from retardation, propagation delay of light between the atoms. In the limit $\ell\rightarrow0$ we have the expansions, keeping the leading terms,
\beq
\Delta(\ell)-\Delta \simeq -\frac{3\gamma}{\ell^3},\,\gamma(\ell) \simeq 2\gamma\,.
\eeq
The shift $\Delta(\ell)-\Delta$ diverges as  $\ell^{-3}$, which clearly reflects the dipole-dipole interactions between the atoms. It is this shift that leads to the decoupling of two closely spaced atoms from the light. On the other hand, the linewidth doubles.

\begin{figure}[h!]
\begin{center}
\includegraphics[width=1.0\columnwidth]{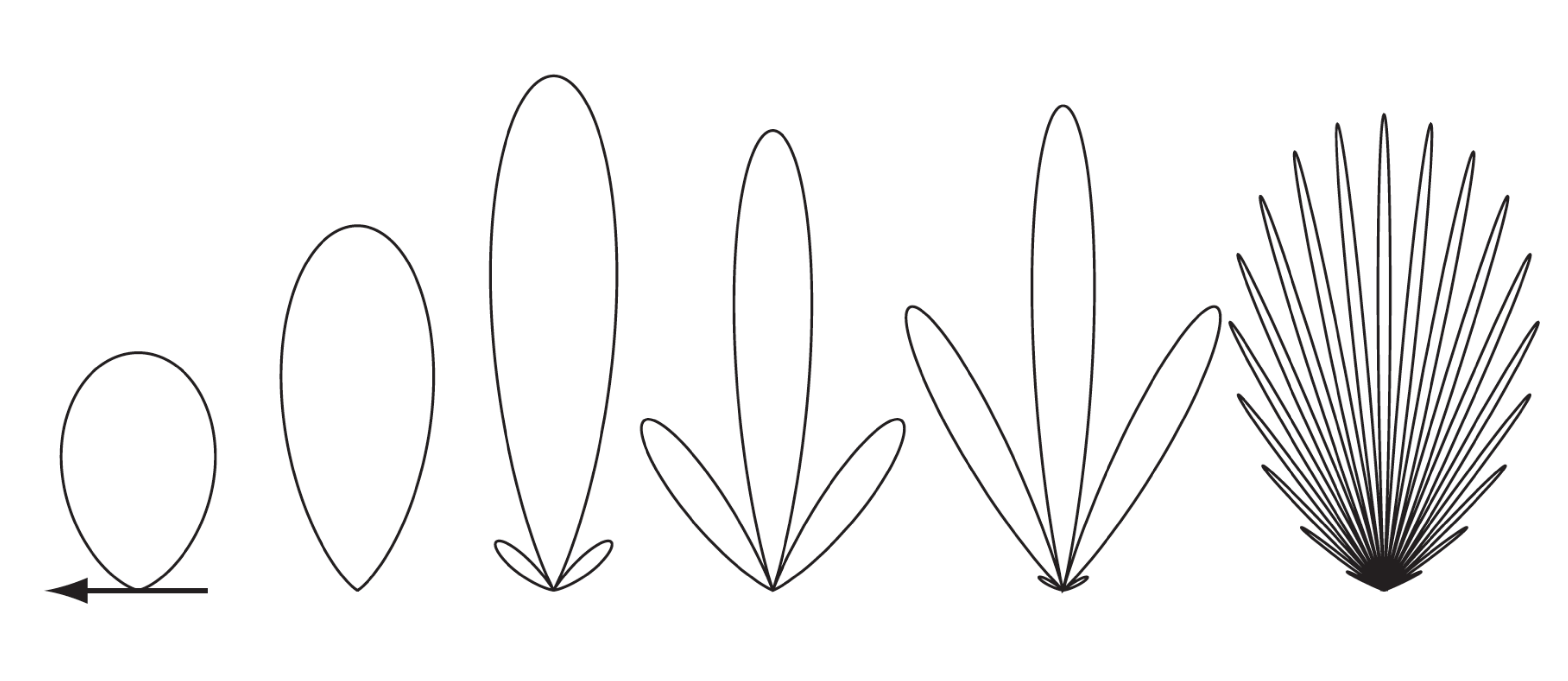}
\end{center}
\caption[Radiation patterns for two dipoles]{Normalized radiation patterns $P(\theta;\ell)\sin\theta$ for two dipoles for the distances between the dipoles $\ell=0,\,\pi,\,2\pi,3\,\pi,4\,\pi$, and $20\pi$ (left to right). These are polar plots for $\theta\in[0,\pi]$, with the common direction of the dipoles and the separation between the dipoles denoted by the arrow in the $\ell=0$ graph.}
\label{TWOATOMS}
\end{figure}

Moving on to the energy flux in the far field and to the radiated power, we find after some tedious mathematics the expressions
\bea
S_S &=& \frac{9 \gamma ^2 |E_0|^2 \cos ^2\!\left[\frac{1}{2} \ell  \cos
   (\theta )\right] \sin ^2(\theta )}{8 \pi  \left[\Delta (\ell )^2+\gamma (\ell )^2\right]r^2},\\
P_S &=&\frac{3 \gamma\,\gamma(\ell)|E_0|^2 }{2[\Delta^2(\ell)+\gamma^2(\ell)]}\,.
\eea
The angular distribution of the radiation, normalized in such a way that $\int_0^\pi d\theta\,\sin\theta\, P(\theta)=1$, reads
\beq
P(\theta;\ell) = \frac{3\gamma}{2\gamma(\ell)}\cos^2\left[\frac{1}{2} \ell  \cos
   (\theta )\right] \sin ^2(\theta )\,.
\eeq
This shows the dipole radiation pattern modulated by the interference of the radiation from the two dipoles; see the demonstration in Fig.~\ref{TWOATOMS}. For $\ell=0$ we have the usual dipole radiation. With increasing $\ell$ the interference first concentrates the radiation more to the $\theta=\pi$ plane perpendicular to the dipoles. With increasing $\ell$ the side lobes grow numerous, and the overall angular distribution pattern rounds out.

As to the radiated power, the difference of the laser frequency from the atomic resonance shifted by the cooperative effects is the true gauge of the detuning. We momentarily assume that this shift of reference point is implicit in the expression of the power in the two-atom radiation $P_S$, and simply replace $\Delta(\ell)\rightarrow\Delta$. On the other hand, if there were no cooperative effects or interference between the radiations from the two dipoles, the total power  would be twice the  power radiated by one dipole under the same driving field. The ratio of the actual two-atom power and the power from two independent dipoles
\beq
C = \frac{P_S(N=2)}{2 P_S(N=1)} = \frac{\gamma(\ell)[\Delta^2+\gamma^2]}{\gamma[\Delta^2+\gamma^2(\ell)]}
\eeq
is a quantitative measure for the effects of the presence of two dipoles. In the limit $\ell\rightarrow\infty$, $\gamma(\ell)\rightarrow\gamma$, and we have $C\rightarrow1$, as expected; the radiated powers from the two dipoles simply add.

The situation is more intriguing in the opposite case $\ell\rightarrow0$ with $\gamma(\ell)\rightarrow 2\gamma$, and the nature of the result depends in an interesting manner on the detuning. With $|\Delta|\gg\gamma$, we have $C\simeq2$, and the two-atom sample radiates twice as much power as two separate atoms would. If $|\Delta|\ll\gamma$, we have $C\simeq\half$. This means that on resonance the two atoms together emit the same power as one atom would.  In fact, in the limit $\ell\rightarrow0$ the radiated power can be expressed at all detunings as
\beq
P_S = \frac{3 |E_0|^2}{4\{[\Delta/2\gamma]^2+1\}}\,,
\eeq
as if we had a single dipole with the dipole moment matrix element equal to $\sqrt2$ times the original dipole moment matrix element, hence the linewidth $2\gamma$. Now, if the two dipoles were at the same place but completely independent, the total induced dipole moment would be twice the dipole moment induced on one atom. But the induced dipole moment is proportional to the square of the dipole moment matrix element, hence the multiplier $\sqrt2$ in the dipole moment matrix element is consistent with the notion of independently radiating atoms. There is a similar $\sqrt2$ in the quantum mechanics of the Dicke states. From the present angle, this factor is classical physics in disguise.

Sufficiently far off resonance the dipoles, even if close to one another, are independent, and each radiates the same amplitude as one dipole would. This means twice the amplitude and four times the power, which is the result we already noted. This is again an interference effect, and has nothing to do with cooperativity. Cooperativity is clearly responsible for the shift of the resonance.

Regarding the modifications of the linewidth, there is some ambiguity. As is well known, on resonance the power radiated from a two-level atom is independent of the dipole matrix element. This may be thought of as a consequence of energy conservation: With increasing dipole moment the atom tends to radiate more, but at the same time the increased radiation damps the resonant response more and these effects exactly balance. Viewed in this way, on resonance even two independent atoms should radiate the same power as one atom, and we might call this an interference effect; four times as much radiation, but also four times as much damping. Nonetheless, we may regard the resonance behavior as cooperative as well. Light from both atoms falls back on both of them, and we have a cooperative radiation reaction that determines the altered damping rate and linewidth. For one thing, the variation of the linewidth $\gamma(\ell)$ with $\ell$ shows that the propagation of light from one atom to the other is involved. The limit $\gamma(\ell)=2\gamma$ for $\ell\rightarrow0$ appears to allow one to think of the resonance linewidth both as an independent-atom phenomenon and as a cooperative phenomenon. This is a remarkable coincidence, if a coincidence it is.

\subsection{What's missing?}\label{TROUBLE}
There are several obvious pieces of physics missing from our picture that may figure in the interpretation of the experiments and simulations alike. We mention a few most notable items here, and amplify as we go along.  An extended and somewhat
complementary account is given in Ref.~\cite{Guerin16b}.

First, there is the interference of the scattered light with the incoming light. This is behind the ``absorption'' of light. Atomic samples that scatter light elastically return all of the light energy back to the light field, and there is no genuine absorption. Instead, the light from the incident driving field, say, a laser beam, and the forward-scattered light interfere destructively. The energy that gets removed from the incident beam is directed elsewhere.

Second, suppose we actually did have a continuous and nonfluctuating distribution of polarization $\bP(\br)$. A typical microscopic model would state that for a continuous density $\rho(\br)$ the polarization is $\bP(\br)=\rho(\br)\bd(\br)$ if the dipole moment of an atom at $\br$ were $\bd(\br)$. The standard method to analyze this situation is to use the  EDPM. It is not an independent-atom theory, but takes into account the effects of the radiation from the atoms on each other in some averaged way. In fact, EDPM is a MFT. Whether one can solve it accurately is another matter, but one can apply general intuition. For instance, a Gaussian cloud might act like a (poor-quality) converging or diverging lens depending on the sign of the dielectric constant, which in turn depends on the sign of the detuning. EDPM is not an exact and possibly not even a quantitatively useful description of the response of an atomic sample to light, but optics-like effects should be expected to be present in the results of both experiments and simulations.

Ordinarily, when one thinks of light propagation through a sample as a standard optics problem, the initial and scattered field are dealt with at the same time. There is the remarkable Ewald-Oseen extinction theorem~\cite{BOR99}, which roughly says that inside a dielectric medium the electric field has a component that cancels the incoming field. That is why the light inside a dielectric medium has the wavelength $\lambda/n$ appropriate for the dielectric constant of the medium $n$, even if from a microscopic standpoint one also concludes that the field inside is the sum of the incoming field and the field scattered from the atoms.

Finally, suppose one solves the light propagation problem numerically using classical-electrodynamics simulations. Since the positions of the atoms are random, there are fluctuations in the scattered radiation, and perforce, in the interference of the scattered and incoming radiation. Smooth radiation patterns as in Fig.~\ref{ANGULAR} are averages over a large number of atomic distributions, but at least over time scales such that the atoms may be regarded as being at standstill there is no such averaging. An individual sample of atomic positions may give a radiation pattern that looks quite ragged. Eventually one has to confront the possibilities of spatial fluctuations in the scattered light, incoherent scattering, and generalizations of incoherent scattering beyond the single-scattering framework.

\section{The slab}\label{THESLAB}

In this paper we study mostly a slab of matter, with the light coming in to a face of the slab at normal incidence. In this case EDPM may be solved exactly in what amounts to a student exercise. The idea is to compare these exact solutions of the MFT with ab-initio numerical simulations.

\subsection{Elementary optics}
\label{ELEMOPTICS}
\begin{figure}
\includegraphics[width=0.6\columnwidth]{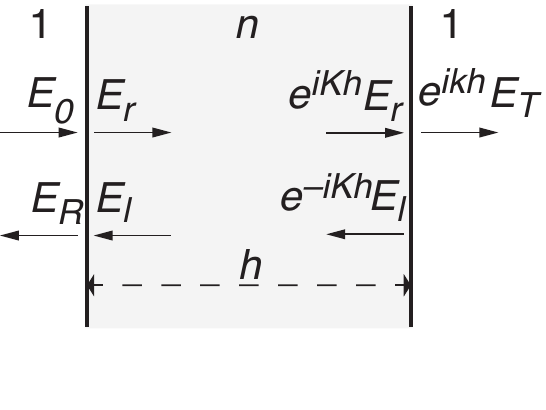}
\caption{Schematic representation of light propagation through a slab of thickness $h$ and refractive index $n$.  }
\label{SLAB}
\end{figure}

Figure~\ref{SLAB} illustrates the standard-optics problem.  The light is coming in from vacuum with the refractive index 1. We denote the refractive index of the medium by $n$, so the wave number inside is $K = n$ ($nk$ in SI units). The incoming light with the reference electric field amplitude at the entrance $E_0$ gets either reflected or transmitted at the entrance face, with the corresponding amplitude reflection and transmission coefficients being $(1-n)/(n+1)$ and $2/(n+1)$~\cite{Jackson,BOR99}. Inside the medium we have two amplitudes, $E_r$ corresponding to the right-going wave with the propagation factor $e^{iKz}$, and the left-going amplitude $E_l$. By matching the incoming, reflected and transmitted waves at the front face we have
\beq
E_r=\frac{2 }{n+1}\,E_0+\frac{1-n}{n+1}\,{E_l}.
\eeq
Similar matching can be made at the exit face,
which leads to the relation between the incoming and transmitted amplitudes
\beq
\frac{E_T}{E_0}=\frac{2ne^{-ih}}{2n\cos nh-i(n^2+1)\sin nh}\,.
\label{ET}
\eeq

To complete the exercise we note that, according to the local-field corrections the effective electric field inside the sample is $E_e = E +\frac{4\pi}{3}P$~\cite{Jackson,BOR99}, the polarization is $P=4\pi\chi E$, where $\chi$ is the susceptibility, and also $P=\rho\alpha E_e$, where $\rho$ is the atom density and
$\alpha$ the polarizability~\eq{DIMLESSALPHA} of an atom. We then have
\beq
\chi=n^2 - 1
 = - \frac{6\pi\rho}{(\delta-\delta_{L})+i},
\label{NSQR}
\eeq
where $\delta_{L}=-2\pi\rho$ is the Lorentz-Lorenz (LL) shift of the resonance.
Simple algebra gives the power transmission coefficient, optical thickness (depth, density), and the conventional absorption coefficient defined as
\beq
T=\left|\frac{E_T}{E_0}\right|^2,\quad D=-\ln T,\quad {\cal A} = 1-T.
\label{TDADEFS}
\eeq

Standard scattering theory says that if light propagates in a medium with density $\rho$ for a distance $h$, for the scattering cross section $\sigma$ the fraction of light energy that makes it through and the corresponding optical thickness are
\beq
T = e^{-\sigma\rho h},\quad D=h\rho\sigma\,.
\label{BEERLAW}
\eeq
This is Beer's law, and the reason why we usually state our results in terms of optical thickness: If Beer's law were valid, the {\em line shape\/} of the optical thickness, e.g., its variation with the tuning of the driving light, would be independent of the thickness of the sample. The physical thickness would simply be a multiplicative factor.

Clearly, the EDPM solution~\eq{ET}-\eq{TDADEFS} cannot agree with Beer's law exactly. There are interesting lessons to be learned from this discrepancy.  Suppose we have an electric field propagating in the $z$ direction, of the form $E(z) ={\cal E}(z)e^{iz}$, where ${\cal E}(z)$ varies little with $z$ over the scale length 1; basically, over the scale of the wavelength. The so-called slowly varying envelope approximation would then say
\beq
\frac{\partial^2}{\partial z^2} E(z) \simeq  i e^{iz}  \frac{\partial}{\partial z}{\cal E}(z).
\eeq
By assuming a dominant propagation direction and a slowly varying electric field amplitude, the wave-equation for the electric field is converted to a first-order differential equation, and Beer's law follows. What gives in a slab is that the dielectric medium is taken to have an abrupt face, so the electric field cannot be expected to vary slowly over the length scale of a wavelength everywhere. Indeed, there is a counterpropagating reflected wave inside the slab, and not just a single direction of propagation. In physical terms, the problems with Beer's law can be attributed to etalon effects, reflections of light from the faces of the dielectric slab.

When we deal with samples that have relevant features of the size of a wavelength like the slab~\cite{Keaveney2012}, or that are of a size comparable to a wavelength~\cite{Jenkins_thermshift,Jennewein_trans}, the usual approximations of optics such as slowly-varying envelope approximation and  paraxial approximation tend to break down, not to mention ray optics. One then has to solve the full Maxwell equations, an onerous requirement even numerically~\cite{DAV11,WRI12,KAH16,Jennewein_trans}. For the slab the standard optics gives an exact solution to Maxwell's equations. The standard-optics result for a slab may also be derived straightforwardly~\cite{Ruostekoski1997b} from Eq.~\eq{MFT}.

It should be noted that we have inserted the local-field correction~\cite{Jackson,BOR99} by hand. Given the specific form of the polarizability, the result is then exactly the LL redshift of the resonance as a function of the density of the sample, $\delta_{L} = -2\pi\rho$. In the standard SI units it would read $\Delta_L=-2\pi(\rho/k^3)\gamma$.

What is the shift of the resonance is a difficult question operationally. Beer's law~\eq{BEERLAW} says that the line shape of optical thickness is a Lorentzian and the position of the resonance can be easily determined. However, in general we find a line shape that is not Lorentzian, and worse, not symmetric about any particular tuning of the driving light. Keeping this in mind, we next discuss the line shift for the EDPM solution of the slab~\eq{ET}-\eq{TDADEFS}.

Let us assume that the line shape is of the form $D=\rho/(K_0 + K_1\rho)$, where $K_0$ and $K_1$ are independent of the density $\rho$. We expand this form in density $\rho$, also expand the EDPM optical thickness from Eqs.~\eq{ET}-\eq{TDADEFS}  in $\rho$, and choose the coefficients $K_0$ and $K_1$ in such a way that up to second order in $\rho$ we have the same expansions. We find that, up to this order in $\rho$, the line shape is Lorentzian, and is shifted from the one-atom resonance by
\beq
s= \delta_{\rm L} +\hbox{$\frac{3}{4}$} |\delta_{\rm L}|\left( 1-\frac{\sin 2h}{2h}\right).
\label{COLLAMB}
\eeq
This is the ``cooperative Lamb shift'' of Friedberg, Hartmann and Manassah~\cite{Friedberg1973}. Here the first term is the LL shift that we put in by hand, and the second, oscillatory, part comes from the etalon effects.
The theoretical result~\eq{COLLAMB} was recently tested experimentally, albeit in a hot gas with moving atoms~\cite{Keaveney2012}, and found to work quite well, apart from a shift between the theory and the observations proportional to the density of the gas. In a cold and dense trapped cloud of atoms an analogous expression was shown to fail~\cite{Jennewein_trans}, but in a very dilute limit analogous expressions, derived from the standard optics, are expected to provide qualitative estimates for the shift even for cold atoms~\cite{Roof16}.

\begin{figure}
\includegraphics[width=1.0\columnwidth]{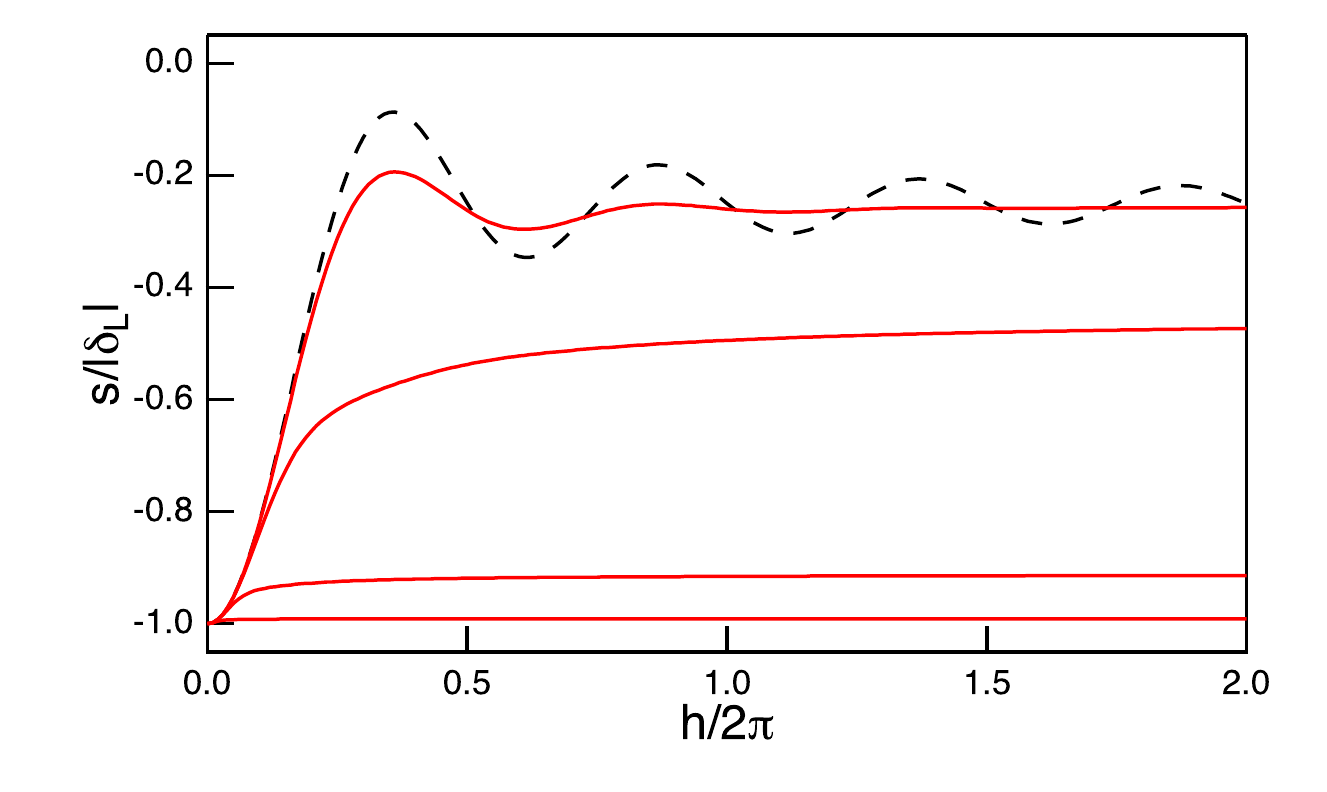}
\caption{Shift of the peak of the resonance line $s$ as a function of the thickness of the sample from MFT. The solid red lines are for the densities $\rho=10$, $1$, $0.1$, and $0.01$, from bottom to top, while the dashed line is the analytical result for an asymptotically low density, Eq.~\eq{COLLAMB}, corresponding to the ``cooperative Lamb shift''. The optical thickness of the sample increases with an increasing density of the atoms, resulting in the deviations of the shift from Eq.~\eq{COLLAMB}, even when the light-induced correlations are not incorporated in the calculation. At high atom densities the MFT resonance shift tends to the LL shift. }
\label{MFTSHIFTS}
\end{figure}

For an arbitrary density we use the maximum of the resonance line as a proxy of the position of the resonance, hence as the shift from the one-atom resonance.  The results obtained numerically from~Eqs.~\eq{ET}-\eq{TDADEFS} are plotted in Fig.~\ref{MFTSHIFTS}. We show the shift in units of the (absolute value) of the LL shift as a function of sample thickness for various densities as solid red lines, and the low-density limit~\eq{COLLAMB} as the dashed black line. At low atom densities the oscillatory etalon effect of the ``cooperative Lamb shift'' is clearly observable. However, as the density increases, the sample becomes optically thicker and the fraction of the light that propagates from one face to the other decreases. Etalon effects are reduced, and the MFT resonance shift tends to the LL shift.

\subsection{Numerical simulations}
\label{NUMSIM}
A slab extending to infinity in the transverse directions would correspond to an infinite number of atoms, an impossibility for numerical analysis. In our simulations we attempt to do the next best thing: We study an atomic sample confined to a circular disk of radius $R$ and thickness $h$, and make the radius as large as practicable. The area of the disk is then $A=\pi R^2$. We put some given number $N$ atoms evenly distributed inside the disk, which gives the number density $\rho=N/(hA)$. We again assume a plane wave of light propagating along the axis of the disk, and denote the direction of propagation by $z$. In keeping with the symmetry of the disk, we take the incoming light to be circularly polarized, so that it again reads
$
\bE_0(\br) = E_0\,e^{iz}\, \hat{\bf e}_+
$.

Simulating the scattered field for an individual sample of atoms and averaging over the samples is straightforward per se. In contrast to the standard-optics solution of Sec.~\ref{ELEMOPTICS}, we do {\em not\/} put in any ad hoc local field corrections or LL shifts. As far as the microscopic model of the dipolar medium is concerned, they simply do not belong there.

\begin{figure}
\includegraphics[width=0.9\columnwidth]{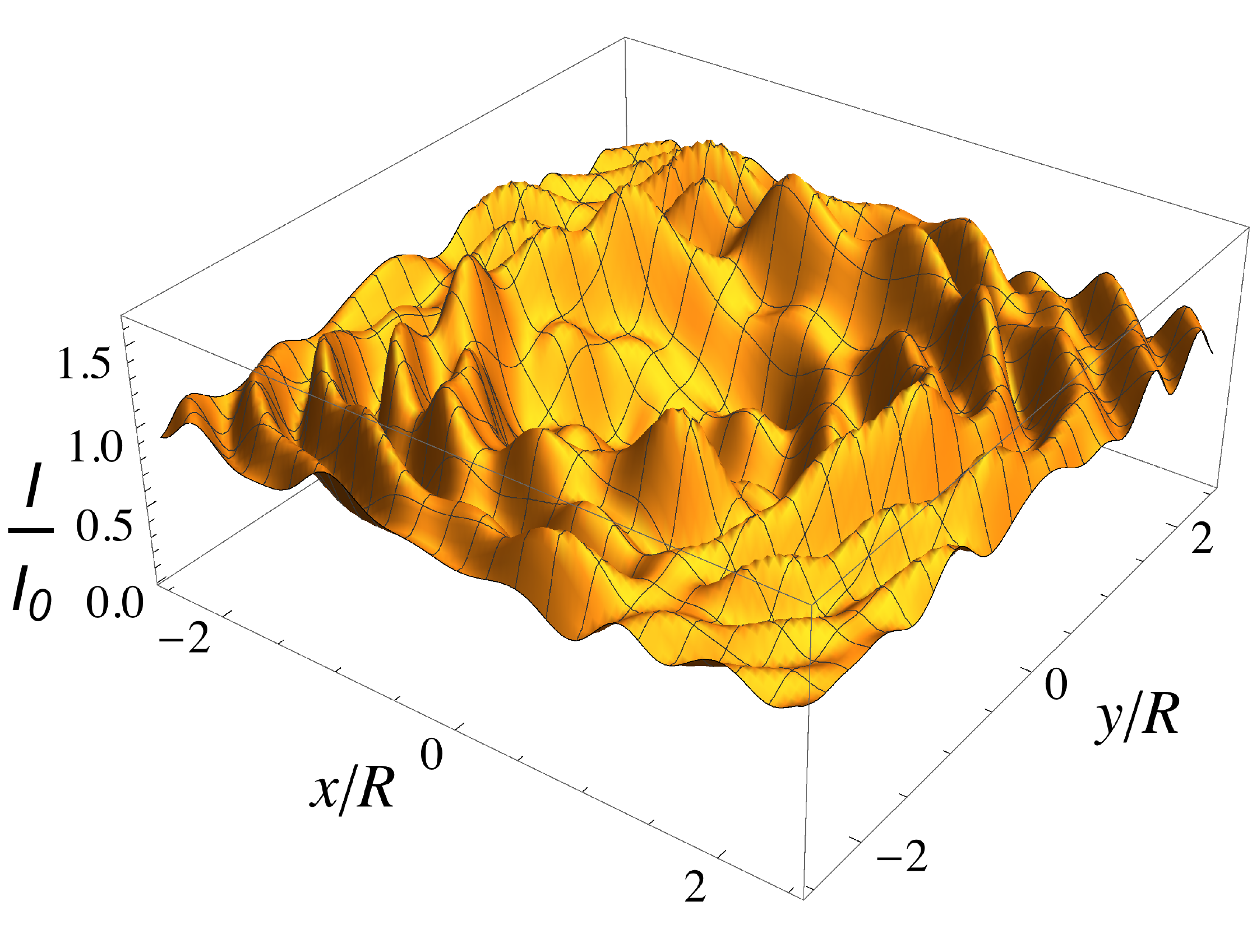}
\caption{Transmitted light intensity at a distance $10\pi$ downstream from (the center of) the disk with thickness $h=1$ and area $A=1024$, and with $N=2048$ atoms inside in fixed random positions.}
\label{DIFFRACTIONFIGURE}
\end{figure}

Here we mostly discuss the transmission of light through the sample. Unfortunately, for the reasons we already touched upon in Sec.~\ref{TROUBLE}, the general simulation scheme we have described previously would not work satisfactorily.  The problem is graphically illustrated in Fig.~\ref{DIFFRACTIONFIGURE}. Here we have a disk with thickness $h=1$ and area $A=1024$, and with $N=2048$ atoms inside in random positions. The figure shows the ratio of the transmitted intensity $I=\frac{1}{8\pi}|\bE_T|^2$ to the incoming intensity $I_0$ in a plane parallel to the disk and at the distance $10\pi$ (five wavelengths) downstream from the center of the disk. One can see the depression at the center, the shadow cast by the disk, but also diffraction rings and fluctuations of the intensity as a function of position. Except for very low atom densities, we cannot handle numerically large enough disks to materially eliminate the diffraction, which would have a large effect on the computed transmission. Besides, the spatial fluctuations of intensity will be present regardless.

There is, however, a shortcut~\cite{CHO12} that appears to  expedite the approach to the limit of large area of the disk enormously. We compute the transmitted intensity outside of the disk as if the atoms only scattered light in the forward direction, and hence call this the forward-scattering approximation. To begin with, let us take a dipole in the $xy$ plane in a disk with radius $R$ centered at the origin, and an observation point in the far field with $x=0$, $y=0$, $z=\xi$ and $\xi\ge1$. We also assume that $R\gg\xi$.  In the argument we take the dipole to be polarized in the $x$ direction, $\bd = d\, \hat{\bf e}_x$, and have it reside somewhere in the disk with the coordinates $\{\varrho \cos\phi, \varrho\sin\phi,0\}$.

In fact, we take the position of the dipole to be random and evenly distributed inside the disk, so that the average of the field at the observation point is
\bea
&&\bar\bE(\xi\hat{\bf e}_z)=\nonumber\\
&& \frac{d}{\pi R^2}\!\!\int_0^R\!\!\! \varrho\, d\varrho\int_{-\pi}^{\pi}\!\!\!d\phi\,{\sf G}_F(-\varrho\cos\phi \,\hat{\bf e}_x\!-\!\varrho\sin\phi\, \hat{\bf e}_y\!+\!\xi\hat{\bf e}_z) \,\hat{\bf e}_x.\nonumber\\
\eea
Here ${\sf G}_F$ retains only the far-field contribution  $\propto 1/r$ to the dipole field propagator $\G$ of Eq.~\eq{DIMLESSG}. We first carry out the angular integral, and in the remaining integral over $\varrho$ make the substitution $x=\sqrt{\xi^2+\varrho^2}$. This results in the electric field
\beq
\bar\bE(\xi\hat{\bf e}_z)=   \frac{d}{R^2}\,\hat{\bf e}_x\int_\xi^{\sqrt{\xi^2+R^2}} dx\,e^{ix}\left(1+\frac{\xi^2}{x^2}\right).
\eeq
This boils down to two integrals. We first have
\beq
\int_\xi^{\sqrt{\xi^2+R^2}} dx\,e^{ix} = i\, e^{i\xi} - i\, e^{i\sqrt{\xi^2+R^2}}\simeq i e^{i\xi}\,.
\eeq
This estimate says that for $R\gg\xi$ the value of the second term oscillates rapidly with a large $R$, and we use its average value 0. The other integral we approximate as
\beq
\int_\xi^{\sqrt{\xi^2+R^2}} dx\,e^{ix} \frac{\xi^2}{x^2}\simeq \int_\xi^{\infty} dx\,e^{ix} \frac{\xi^2}{x^2}
\rightarrow ie^{i\xi}\,
\eeq
where the second form is found numerically for the limit $\xi\gg1$. The same argument could just as well be made for the $y$ polarization of the dipole, so for a dipole $\bf d$ in the $xy$ plane we simply have
\beq
\bE_T(\xi\hat{\bf e}_z)= 2i\,\frac{{\bf d}}{R^2}  e^{i\xi}.
\eeq
Here $e^{i\xi}$ is a phase-matching factor as dictated by the driving plane wave of light.

Average over the circle with the radius $R$ eliminates the longitudinal component of the dipolar field, but it may be present in the field of a dipole that is not on the axis of the circle. In that case we remove the longitudinal component by hand. Given the incoming plane wave of light $\bE_0 e^{i z}$, we therefore write the total transmitted light as a sum over the dipoles at their positions $\br_k$ as
\beq
\bE_T(\br) = \bE_0 e^{i\xi} + \frac{2i}{R^2}\sum_k [\bd(\br_k)-\hat{\bf e}_z \cdot\bd(\br_k)\,\hat{\bf e}_z] e^{i(\xi-z_k)}\,.
\label{TRLIGHT}
\eeq
This is the same prescription as given in~\cite{CHO12}, albeit in our system of units.

We have compared the transmission coefficient calculated from Eq.~\eq{TRLIGHT} with the analytically known result for one atom in the disk obtained from scattering theory. Suppose the light is on resonance so that the scattering cross section is $6\pi$, then to the leading order the analytical approximation of the absorption coefficient is ${\cal A}=1-6\pi/A$. For a disk with the area $A=256$, the difference from this prediction and the forward-scattering approximation is about $2\%$, and the difference decreases inversely proportionally to the area of the disk. In the limit of a dilute sample the forward-scattering approximation also reproduces the MFT results, the main difference being in the shift of the resonance. We see this even if the disk is so thick that most of the light is absorbed. Given Fig.~\ref{DIFFRACTIONFIGURE} and the ugly approximations in the derivation, Eq.~\eq{TRLIGHT}  reproduces the MFT results amazingly well when MFT is expected to be valid.

We have interspersed  test cases among our simulations where we have increased the area of the disk and looked for convergence of the transmission coefficient. To verify the convergence has proven to be exceedingly expensive in computer time, but we obtain order-of-magnitude estimates of the simulation error due to the finite size of the disk. We occasionally quote them with our results. These truncation errors are usually the largest known numerical errors in our computations, surpassing the statistical fluctuations that result from the necessarily finite number of samples used in the averaging over the atomic positions.

The forward-scattering approximation together with the increasing size of the disk can evidently be used to mitigate the complications due to the optics of the finite-size disk. However, the forward-scattering approximation also removes the spatial fluctuations from the transmitted light. To quantify the fluctuations we momentarily discuss the reradiated dipolar field only, the sum on the right-hand side of Eq.~(\ref{FIELDFROMDIPOLES}). We imagine placing a probe disk of the same radius $R$ as the simulation sample, at the distance $R^2/2$ downstream from the atomic sample. This distance is analogous to the Rayleigh range, where the light radiated by the atomic sample starts to transition from the near-field form of a beam of light to the far-field form of a cone with a constant opening angle, and as such gives a natural place where to observe the scattered field. We integrate the component of the electric field with the same polarization $\hat{\bf e}_+$ as the incoming beam over the probe disk, and study fluctuations of the integrated field over the atomic samples. For a disk with $h=1,\,A=1024,\,N=1024$, and on resonance, the fluctuations are about 6\%.

For an infinitely large disk absorption would correspond to the interference of the incoming and scattered light. In the case when the optics of the disk has a significant effect,  the interference between the scattered field and a field with the wavefront matched to the diffraction pattern of a disk-shape aperture might lead to a more meaningful measure of absorption, but the diffraction pattern is difficult to calculate accurately as this would require solving full vectorial Maxwell's equations. Instead, our studies of fluctuations give us an indirect estimate of the limitations of the calculations of the absorption coefficient as we have done them in this paper: No matter what wave front, in our particular example a random residual field of about $6/100$ of the original field amplitude must remain after the incoming and scattered fields have canceled each other to the maximum extent allowed by optics. A fraction of the intensity of about $(6/100)^2$ invariably gets through as a result of the fluctuations. Such fluctuations are not present in EDPM (here we always assume that EDPM refers to a static, continuous medium). We might conceivably think of the transmission of the random component of the field as diffusion of light through a sample of randomly spaced scatterers~\cite{Chandraskhar1960,Ishimaru1978,Rossum}. In our example a meaningful comparison with the MFT would only be possibly at optical thicknesses of less than $D\sim-\ln (6/100)^2 \sim 6$. From this kind of arguments we surmise that diffusion of light does not materially affect the conclusions of the present paper.

\section{Results for the slab}
\label{RESULTS}
\begin{figure}
\vspace{-12pt}
\center\includegraphics[width=0.85\columnwidth]{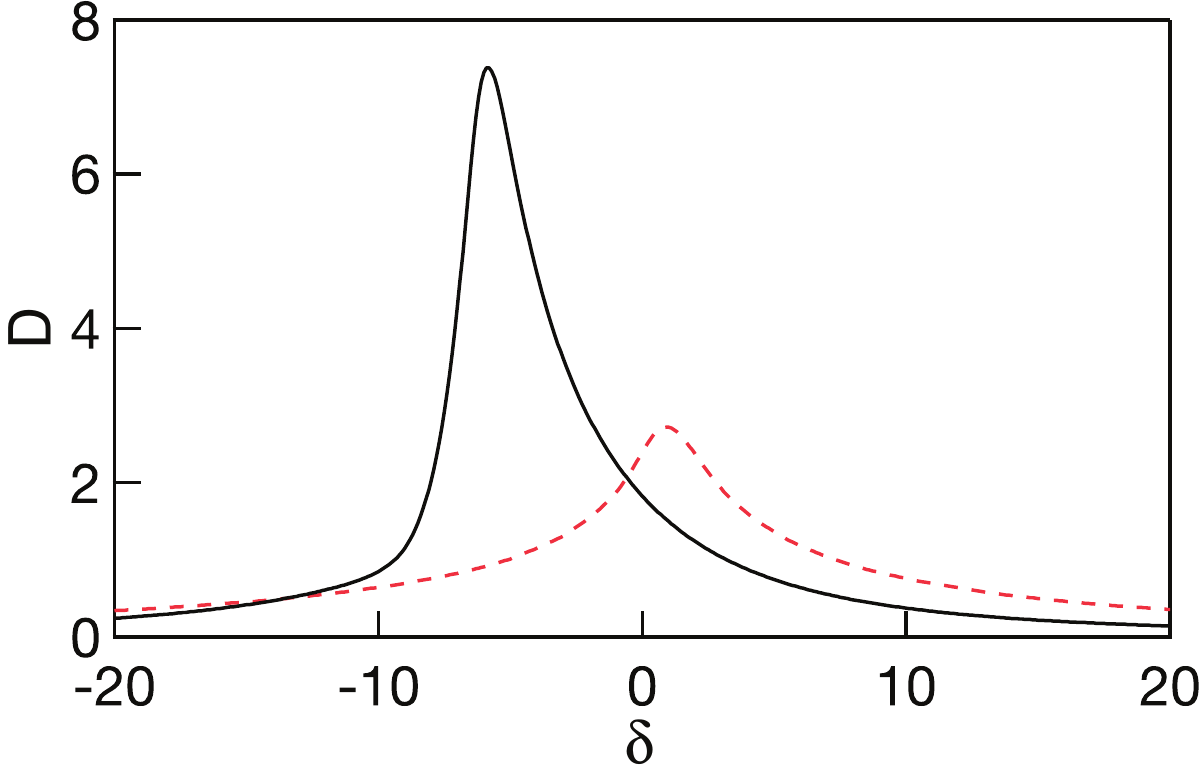}
\vspace{-12pt}
\caption{Optical thickness  of a slab of matter as a function of light-atom detuning from both standard optics (solid black line) and from numerical simulations (red dashed line). The results are for the sample density $\rho = 1$ and slab thickness $h=1$. The truncation error in the numerical computations due to the finite area $A = 1024$ of the disk-shape sample, about 5\%, is irrelevant to this comparison.}
\label{SLABTRANSMISSION}
\end{figure}

Our main qualitative result is displayed in Fig.~\ref{SLABTRANSMISSION} that compares the optical thickness from the MFT and from numerical simulations for a dense ($\rho=1$) slab. Even if one discounts the LL shift that was put in by hand to the MFT anyway, there is a large difference between the curves. Our interpretation is that the standard EDPM fails. This is not hard to fathom: EPMD is a MFT for the light-mediated interactions between the dipoles, and as garden variety MFTs do, it goes bad when the interactions between the atoms increase; here, as a result of density.

\begin{figure}
\vspace{-12pt}
\center\includegraphics[width=0.85\columnwidth]{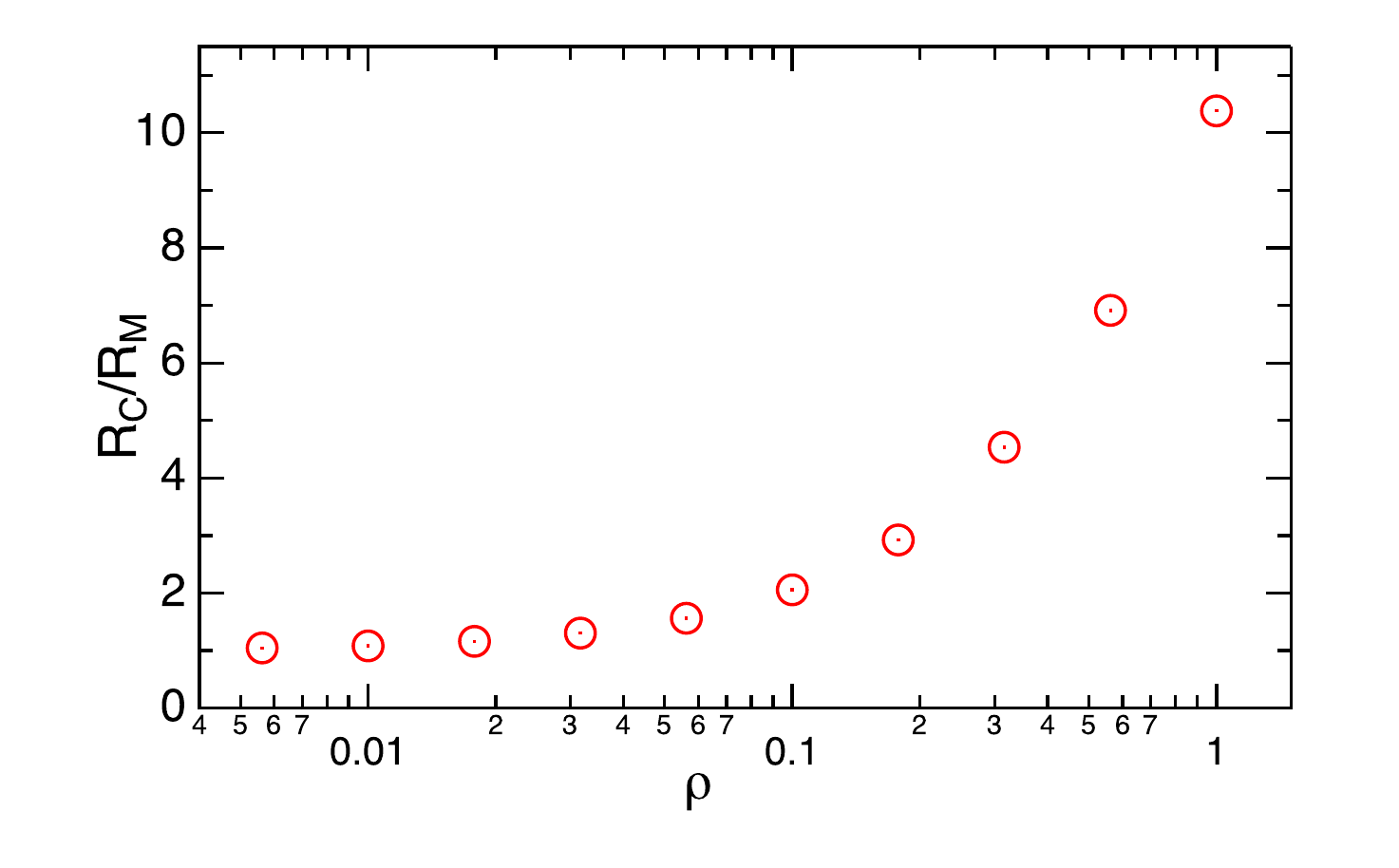}
\vspace{-6pt}
\caption{Ratio of the absorption coefficient from numerical computations, ${\cal A}_C$, and from MFT, ${\cal A}_M$, as a function of the density of the slab with the thickness $h=\pi$. The positive detunings are chosen in such a way that the MFT gives ${\cal A}_M =0.01$ for each density, an optically thin sample. The numbers of atoms, evidently integers, are chose so that for a given density the area of the disk is as close to $A=2000$ as possible.}
\label{SCATTERING}
\end{figure}

Our next question is, what kind of densities are needed for notable discrepancies between the MFT and the simulations? For a demonstration, we have developed the following scheme. We first fix the thickness of the slab at $h=\pi$. For various values of density $\rho$ we then find from the MFT exactly, numerically, the corresponding positive detuning $\delta(\rho)$ for which the absorption coefficient is ${\cal A}_M=0.01$. There is a corresponding negative detuning such that the absorption is the same, not precisely the negative of the positive detuning because the line shape is not symmetric about $\delta=0$, but we arbitrarily pick the positive detunings. We also compute  the absorption coefficients ${\cal A}_C$ from the simulations for the same densities $\rho$ and detunings $\delta(\rho)$. In Fig.~\ref{SCATTERING} we plot the ratio ${\cal A}_C/{\cal A}_M$, the cooperative enhancement of absorption, for a number of sample densities. The enhancement exceeds two already for a sample as dilute as $\rho=0.1$, with $\delta(\rho)\,=\,24.2\,$. The message here is that, depending on what the experiment might be, significant deviations from MFT may be found at unexpectedly low densities even for the off-resonance case.

The results of Fig.~\ref{SCATTERING} also tally with the simulation results in Fig.~\ref{SLABTRANSMISSION}. The detunings are large in~Fig.~\ref{SCATTERING}, about $\delta\sim80$ for the case of $\rho=1$, and it is evident from Fig.~\ref{SLABTRANSMISSION} that this far in the wings the MFT understates optical thickness and absorption.

There are aspect in Fig.~\ref{SCATTERING} that also bear on the interpretation of our simulations. The overall absorption in the figure varies between 0.01 and 0.1 depending on density, so that between 90\% and 99\% of the light gets through. The sample is optically thin, the thickness being at most 1/10 of the ``mean free path'' of a photon. The deviations from MFT evidently cannot be attributed to diffuse scattering or radiation trapping~\cite{Chandraskhar1960,Ishimaru1978,Rossum,BOU13,Guerin16b}. The area of the disk was also large, approximately $A=2000$, and we looked at the convergence of the results with the disk area extensively. Even at its worst, at the highest density $\rho=1$ in the figure, the relative error in the results should be no more than on the order of 5\%. The finite size of the disk should not be a major contributor to the deviations of the simulations from MFT either.

The sample in Fig.~\ref{SCATTERING} is optically thin because of the large detuning. For the MFT, Eqs.~\eq{ET}-\eq{TDADEFS}, an expansion in $1/\delta$ appropriate to the limit of a large detuning $\delta$ gives
\beq
{\cal A}_M = \frac{6\pi\rho h}{\delta^2}\left[1+\frac{3|\delta_{\rm L}|}{8h}(1-\cos 2h) \right] + {\cal O}\left(\frac{1}{\delta^3}\right).
\label{OFFRESABS}
\eeq
The factor in front is the absorption coefficient of independent atoms far-off resonance when the cross section for photon scattering $\simeq{6\pi}/\delta^2$ is asymptotically small. However, even in MFT and far-off resonance, the sample does not behave like a collection of independent radiators. The reason is etalon effects, reflections of light from the faces of the slab: For a large detuning the refractive index $n$ is close to one and deviations from unity scale like $\rho/\delta$. A fraction of light $\simeq 1-n\propto \rho/\delta$ that made it through the slab, almost all of it for a large $\delta$, is reflected from the back face, almost all of the reflected light propagates to the front face, gets reflected again with a reflection coefficient $\propto\rho/\delta$, and finally interferes with the light that goes straight through. That is the reason for the two terms in Eq.~\eq{OFFRESABS} proportional to $(\rho h/\delta^2)$ (light straight through) and $(\rho/\delta)^2$ (interference).

Aside from pointing out that run-of-the-mill optics can pop up in quite unexpected places, we use Fig.~\ref{SCATTERING} and Eq.~\eq{OFFRESABS} as a springboard for dimensional analysis. Our hypothesis is that there are two in principle independent dimensionless density parameters in the light-propagation problem, the on-resonance optical thickness $6\pi\rho h$ ($6\pi\rho h k^{-2}$ in terms of full dimensional quantities) that is a MFT parameter, and the plain $\rho$ ($\rho k^{-3}$) that governs the role of dipole-dipole interactions beyond MFT. In several recent experiments~\cite{wilkowski2,Guerin_subr16,Ye2016,Araujo16} the on-resonance optical thickness has proven to be the dimensionless parameter that governs the density dependence of the results. In contrast, our interpretation is that the nontrivial results  in Fig.~\ref{SCATTERING} are attributed to the beyond-MFT parameter $\rho$.

Figure~\ref{SCATTERING} does not separate the two dimensionless quantities cleanly as we kept the MFT absorption coefficient constant, not the on-resonance optical thickness. It also seems that optics, in this case due to the reflection of light from the surfaces of the slab, inevitably imposes some ambiguity in the interpretation of the results. Here we attempted to minimize the effects of the optics by choosing the sample thickness $h=\pi$, whereupon the etalon-effect term in Eq.~\eq{OFFRESABS} vanishes.

\begin{figure}
\center\includegraphics[width=1.0\columnwidth]{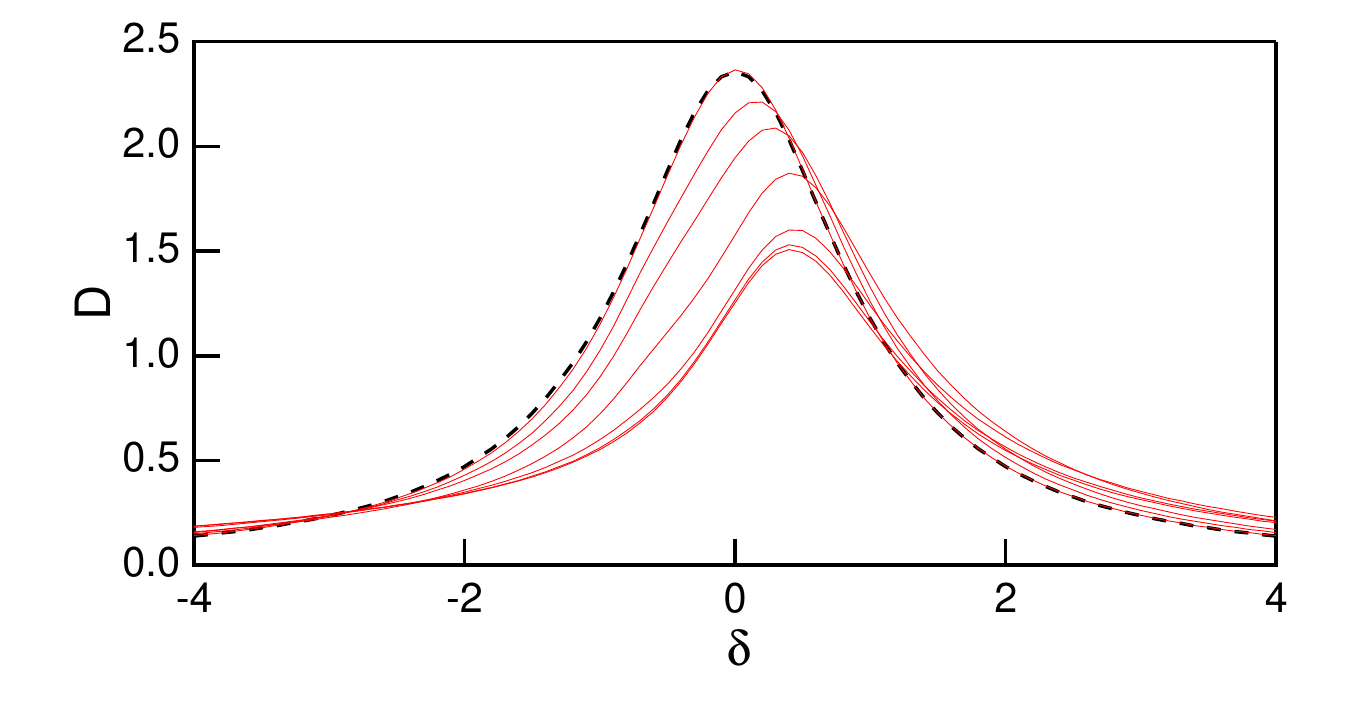}
\caption{Optical thickness $D$ as a function of detuning $\delta$ for varying sample thicknesses $h=0.25$, 0.5, 1, 2, 4, 8, and 16 (red curves from bottom to top). In these figures the area density is kept constant at $N/A = 0.125$, so that the density correspondingly varies from $\rho=0.5$ to $\rho=0.0078125$ by factors of 0.5. The dashed black line is the prediction from Beer's law, which only depends on area density.}
\label{THICKNESSVARIATION}
\end{figure}

We next describe a numerical experiment in which we literally keep the on-resonance optical thickness constant and vary the density. The basic idea is to keep the area of the disk ($A=4096$) and atom number ($N=512$) constant while varying the thickness $h$. The resulting optical thickness as a function of detuning from the independent-atom scattering theory~\eq{INDATSC} would be
\beq
D = \frac{6\pi N}{A(1+\delta)^2};
\label{ANAPRED}
\eeq
$D=2.4$ on resonance with $\delta=0$, which means less than 10\% power transmission. In practice we had to deviate from this ideal to keep the error due to the finite area of the disk somewhat under control, so that at the high end of the thickness range we finished with $N=2048$, $A=16384$ and $h=16$, but of course without altering the preset optical thickness. The resulting absorption line shapes, optical thickness vs.\ detuning, are shown in Fig.~\ref{THICKNESSVARIATION} for a range of thicknesses varying from $h=0.25$ to $h=16$ in multiples of two (solid red lines from bottom to top), corresponding to the densities ranging from $\rho=0.25$ to $\rho=0.0078125$ decreasing by factors of $0.5$. Also shown is the corresponding independent-atom prediction~\eq{ANAPRED} (dashed black line), which basically differs from the lowest-density simulation graph by a very small shift of the resonance frequency.

As before, there are etalon effects that affect the results, although the optical thickness should depress  them some as the light reflected from the back face of the slab should complete a back-and-forth trip before interfering with the light that gets through on the first try. This caveat notwithstanding, we see substantial effects of the varying density on the absorption line shapes. The dipole-dipole interactions make a noticeable difference already at the density $\rho=0.015625$, the second red curve from the top.

There are two other observations to be made here. First, at higher densities the resonance shifts as large as in Fig.~\ref{MFTSHIFTS} should be plain visible to the eye in Fig.~\ref{THICKNESSVARIATION}, and the resonances should move to the red. There are visible density shifts of the resonance alright, but much smaller than one would surmise from Fig.~\ref{MFTSHIFTS} and with the opposite sign. The line shifts large enough that one can plainly see them are not compatible with the MFT, especially not so when the LL shift is included in the MFT. Second, one should note that the graphs for the thicknesses $h=0.25$, $h=0.5$ and $h=1$ are close to one another; the first two are hard to distinguish at all. This holds true even though the density increases by a factor of two for each curve. Our interpretation is that at decreasing thicknesses, here evidently below $h=1$, the physics must eventually become two-dimensional and then it is the area density $N/A$ instead of the volume density $\rho=N/(Ah)$ that governs the scaling. The area density, of course, is the same for all curves in this figure. We have yet another complication to take into account when interpreting the numerical simulations~\cite{CHO12}.

Even though the ``cooperative Lamb shift''~\eq{COLLAMB} was seen in experiments with dense hot gases~\cite{Keaveney2012}, our simulations for dense cold gases have produced no comparable shifts, and no sign of the LL shift either~\cite{Javanainen2014a}. The line shifts from our simulations simply do not agree with the standard expectations. We next discuss two of our quantitative studies of the line shift from Refs.~\cite{Javanainen2014a} and~\cite{JavanainenMFT} in added detail.

\begin{figure}
\vspace{-12pt}
\center\includegraphics[width=0.9\columnwidth]{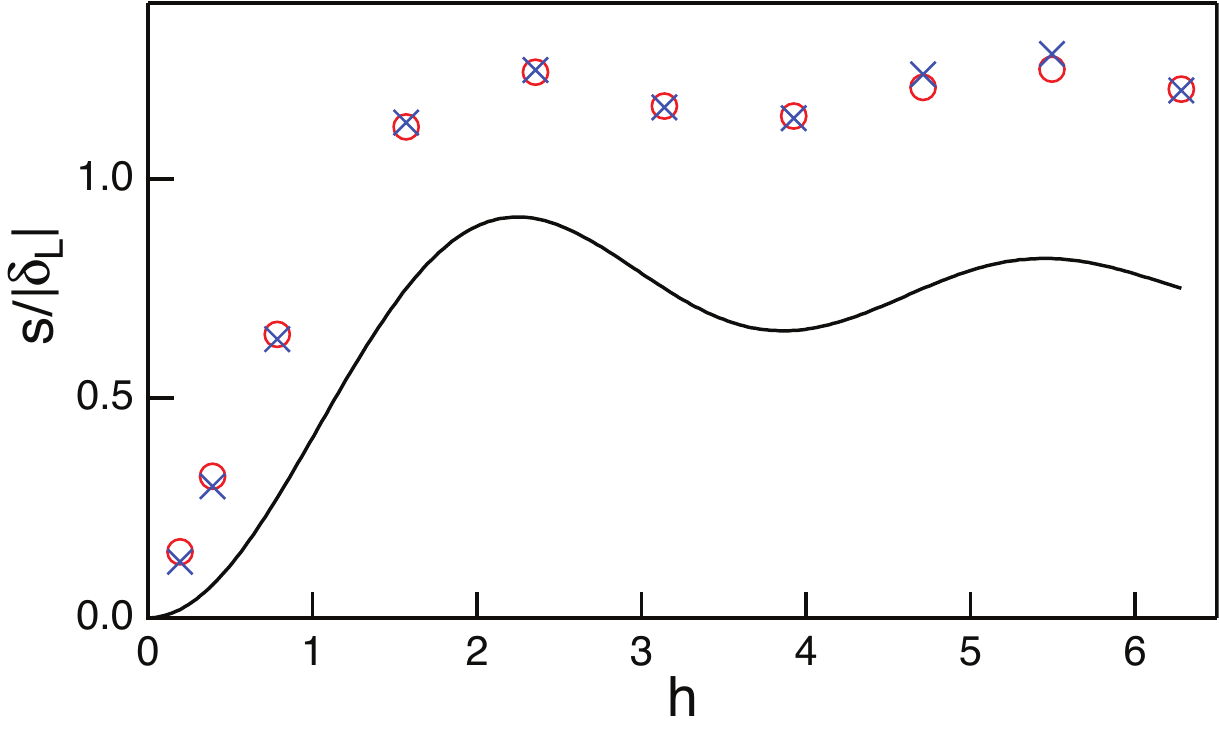}
\vspace{-12pt}

\caption{Shift of the resonance $s$ as a function of the sample thickness $h$, for two densities $\rho=0.01$ (circles) and $\rho=0.005$ (crosses). Also shown is the prediction~\eq{COLLAMB}, shifted up by $|\delta_{\rm L}|$ for easier comparison (solid line).}
\label{LOWRHOSHIFTS}
\end{figure}

The ``cooperative Lamb shift''~\eq{COLLAMB}, in our view, is an etalon effect calculated analytically for the limit of asymptotically low atom density. We therefore study the line shifts as a function of sample thickness $h$ at low atom densities, $\rho=0.01$ and $\rho=0.005$. For the latter the LL shift is 3\% of the natural linewidth, so the position of the resonance has to be found accurately. Fortunately, for such low densities the resonance lines $D(\delta)$ are well approximated by Lorentzians, so we can fit them with a Lorentzian with an adjustable center (and width). We did the fits over the detuning range $\delta\in[-2,2]$, and in a few explicit tests found that the resonance positions for data from different simulation runs were reproducible on a few-percent level. The results are shown in Fig.~\ref{LOWRHOSHIFTS}. Also shown is the prediction~\eq{COLLAMB}, albeit shifted up by the absolute value of the LL shift. The maximum optical thickness in these simulations was $D\simeq1$.

The oscillatory dependence of the line position on the thickness of the sample is evident, but with two caveats. First, the numerical results track the oscillations in the theory fairly well, but only at thicknesses comparable to or larger than $h=1$. Below that, the approximately constant difference between numerical results and the shifted analytical results shrinks to zero, as the numerically computed shifts clearly tend to 0 for $h\rightarrow0$. We attribute the thin-sample behavior to transition from 3D physics to 2D physics, as already discussed in conjunction with Fig.~\ref{THICKNESSVARIATION}. For a fixed density $\rho$, the area density $h\rho$ tends to zero as $h\rightarrow0$, and since in 2D physics it is the area density that counts, the density shift of the line tends to zero as well. Second, there is a large additive constant in the shifts compared to the MFT prediction; recall that we have already removed the LL shift from the theory curve in Fig.~\ref{LOWRHOSHIFTS}. It appears that even in the limit of low atom density there is little quantitative validity to the LL shift.

On the other hand, the experiments that found a variation of the line shift in accordance with the ``cooperative Lamb shift'' and/or the LL shift~\cite{MAK91,Keaveney2012} were carried out in hot atomic vapors in which the atoms move at thermal speeds, and also collide. We have coded classical-electrodynamics simulations of moving atoms, but the convergence of the results is not yet adequately under control. We therefore adopt a shortcut. Namely, to the lowest order of approximation the motion of the atoms causes Doppler shifts, and as a result the resonance frequencies of the atoms appear to have a corresponding random distribution. We simply add such inhomogeneous broadening to our simulations: While generating a random position for each atom, we also add a random shift to the resonance frequency drawn from a Gaussian distribution with zero average and the rms width $\Omega=100$. This in fact is a reasonable estimate for the D lines in near-room temperature alkali vapors.

The result is a spectrum $D(\delta)$ that under a casual inspection looks like a Gaussian with the rms width $\Omega$. The assignment is to find the center of the resonance line. We resort to a standard method in experimental spectroscopy: We define what is known as the Voigt profile, convolution of a Lorentzian  (width $\Gamma$, unit height) and Gaussian with a width $\Omega$,
\bea
V(\delta,\Gamma,\Omega) &=&  \frac{1}{\sqrt{2\pi}\,\Omega} \int d\zeta\, e^{-\frac{\zeta^2}{2\Omega^2}}\frac{\Gamma^2}{(\delta+\zeta)^2+\Gamma^2}\\
&=& \sqrt{\frac{\pi}{2}}\,\frac{\Gamma}{\Omega}\,\Re
\left[
e^{\frac{\Gamma-i\delta}{2\Omega^2}}{\rm erfc}\left(
\frac{\Gamma-i\delta}{\sqrt2 \,\Omega}
\right)
\right],
\eea
where erfc is the complement of the error function as defined, say, in Mathematica, and fit the observed line shape $D(\delta)$ to a Voigt profile $H\,V(\delta-s,\Gamma,\Omega)$. Here we regard the overall height $H$, the shift $s$, and the widths of the Lorentzian $\Gamma$ and of the Gaussian $\Omega$ all as adjustable parameters.

\begin{figure}
\includegraphics[width=1.0\columnwidth]{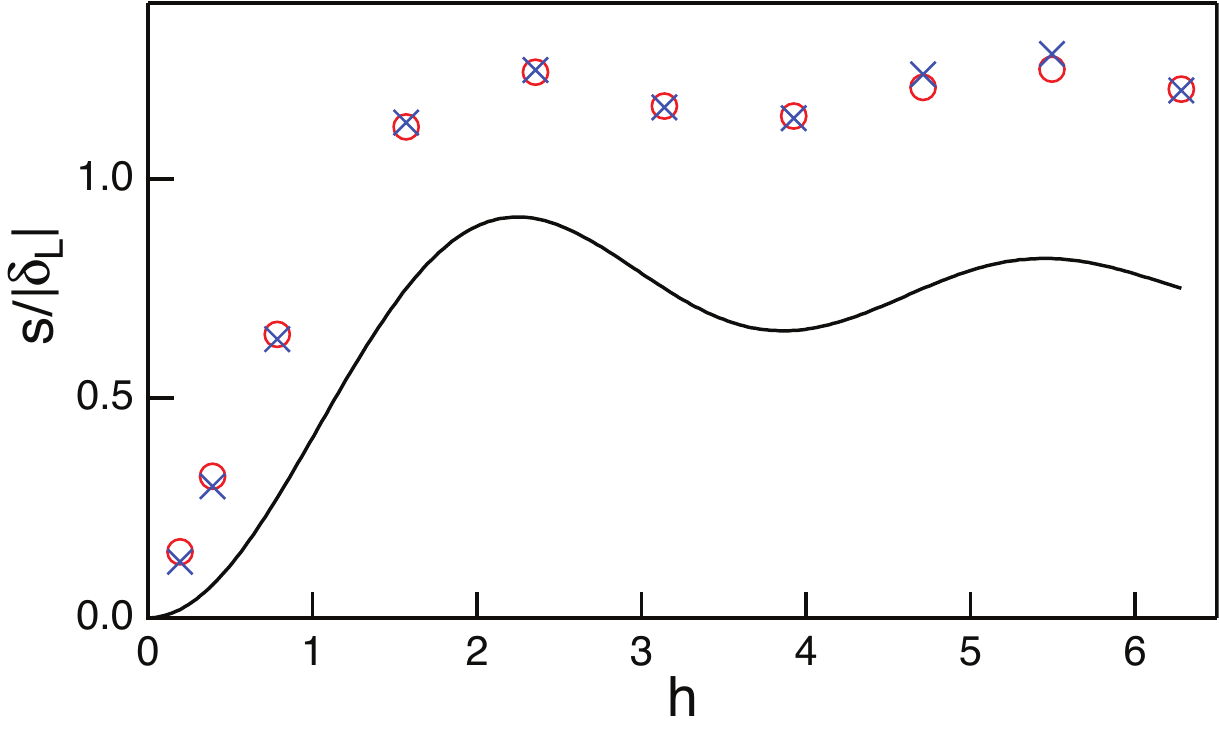}
\caption{The shift of the absorption line $s$ plotted as a function of the thickness of the sample $h$ as solid circles for the gas density $\rho=1$ and inhomogeneous broadening of $\Omega=100$.  Also shown as a solid line is the ``cooperative Lamb shift'', Eq.~\eq{COLLAMB}, and as a dashed line a vertically translated version of Eq.~\eq{COLLAMB} fitted to the numerical data points~with~$hk\ge1$.}
\label{INHLINE}
\end{figure}

Figure~\ref{INHLINE} presents the shift of resonance line $s$ as a function of the sample thickness for a disk with the density $\rho=1$, given inhomogeneous broadening with the rms value $\Omega=100$. The dots are from numerical simulations, the sizes being comparable to our estimate of the statistical errors. The solid line is the ``cooperative Lamb shift'' as from Eq.~\eq{COLLAMB}, and the dashed line is a vertically translated version of Eq.~\eq{COLLAMB} that gives the best fit to the numerical data points with $h\ge 1$. The fits of the simulated line shapes to the Voigt profile turn out to be excellent, and in our examples we obtain reproducible results for the shifts $s$ that are on the order 1\% of the width of the Gaussian $\Omega$. However, practical constraints forced us to a rather small disk area of $A=256$, which contributes an unknown truncation error. Again, the maximum optical thickness in the samples used to prepare the figure was on the order of $D\simeq1$.

We once more attribute the exceptional behavior of thin samples, $h\lesssim1$, to the transition from 3D to 2D physics. Other than this, the simulation results are quite close to the prediction Eq.~\eq{COLLAMB}, the deviation being about $0.4\,|\delta_L|$. There was a similar ``collision shift'' in the recent experiments~\cite{Keaveney2012} that verified the prediction~\eq{COLLAMB}, so the agreements of our numerical experiments and real laboratory experiments with the theory~\eq{COLLAMB} are on a similar footing.

\begin{figure}
\includegraphics[width=1.0\columnwidth]{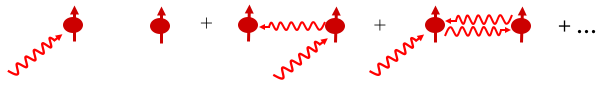}
\caption{Schematic illustration for the excitation of two coupled atomic dipoles by light. In the first term an incident light drives the left atom. The second term represents the excitation of the left atom by light scattered from the right atom
that is excited by the incident field. In the third term the left atom is excited by light that is then scattered back to the left atom via the other atom. Each subsequent term in the series includes an increasing number of scattering processes between the atoms.}
\label{fig:diag}
\end{figure}

The effect of inhomogeneous broadening is to modify the optical response by emphasizing the mean-field phenomenology via the suppression of light-induced correlations between the atoms.
The basic principle is simple to understand: with increasing inhomogeneous broadening the atoms are farther away from resonance with the light that mediates the interactions between the atoms.
We can illustrate the interplay between the inhomogeneous broadening and light-mediated interactions by a simple two-atom example~\cite{Javanainen2014a}. The atoms 1 and 2 are assumed to have different resonance frequencies, hence
different polarizabilities $\alpha_1$ and $\alpha_2$. The field amplitude at the atom 2 is then the sum of the incident field amplitude and the field scattered by the atom 1. Formally, we can write it as
\vspace{-6pt}
\bea
&&\bE(\br_2) = \frac{{\bf E}_0(\br_2) + \alpha_1\G {\bf E}_0(\br_1)}{1-\alpha_1\alpha_2 \G\G}\nonumber\\
&&= {\bf E}_0(\br_2)  + \alpha_1\G {\bf E}_0(\br_1) + \alpha_1\alpha_2 \G\G {\bf E}_0(\br_2) + \ldots\,.
\vspace{-6pt}\eea
The operator expression in the denominator in the first line is expanded in a power series,
as illustrated in Fig.~\ref{fig:diag}. The first term is the free field on atom 2; in the second term the free field excites atom 1, which sends its dipolar field back on atom 2; in the third term the free field excites atom 2, which sends a dipolar field to excite atom 1, which sends a dipolar field back on atom 2. Further terms in the expansion come out the same way reflecting repeated photon exchanges between the atoms. The last term shown is also the first example of a recurrent scattering process in which a light wave interacts more than once with the same atom.

Let us now regard atom 2 as the spectator and imagine averaging over the position of atom 1. This operation faces major mathematical obstacles because of the divergence of $\G(\br_1,\br_2)$, but we do not attempt to sort them out because these problems are evidently similar for homogeneously and inhomogeneously broadened samples. Next add the inhomogeneous broadening $\omega_D$. To the order of magnitude, averaging over the resonant frequencies suppresses the polarizability by a factor of $\gamma/\omega_D$. Thus, the first nontrivial term in the expansion corresponding the mean-field polarization gets suppressed by this small factor, and the higher terms by higher powers of the small quantity $\gamma/\omega_D$. Qualitatively, repeated photon exchanges are de-emphasized because in such processes both the emitter and the absorber are off resonance.

Our numerical simulations confirm analogous behavior in many-atom ensembles. In fact, when the inhomogeneous broadening $\omega_D$ exceeds the resonance linewidth $\gamma$ of the atoms, the results begin to approach the mean-field phenomenology of standard optics, indicating that macroscopic EDPM is an emergent theory, resulting from the suppression of light-induced correlations. The same effect was demonstrated experimentally in the case of fluorescence where the resonance shifts of a cold, dense gas of atoms substantially differed from those predicted for thermal atomic ensembles~\cite{Jenkins_thermshift}. Both experimental observations and numerical simulations revealed the absence of any notable shift in cold trapped atomic ensemble. However, introducing inhomogeneous broadening in the simulations restored a large value for the shift. 

The suppression of light-mediated interactions by inhomogeneous broadening is a generic effect in  coupled resonant emitter systems. For instance, electromagnetic interactions between solid-state radiators, such as plasmonic circuit resonators, may be described by analogous coupled-dipole model simulations~\cite{JenkinsLongPRB}. Inhomogeneous broadening in such a system can result, e.g., from fabrication imperfections, and have been shown to notably suppress strong radiative interactions between the resonators~\cite{JenkinsRuostekoskiPRB2012b}.

\begin{figure}
\includegraphics[width=0.9\columnwidth]{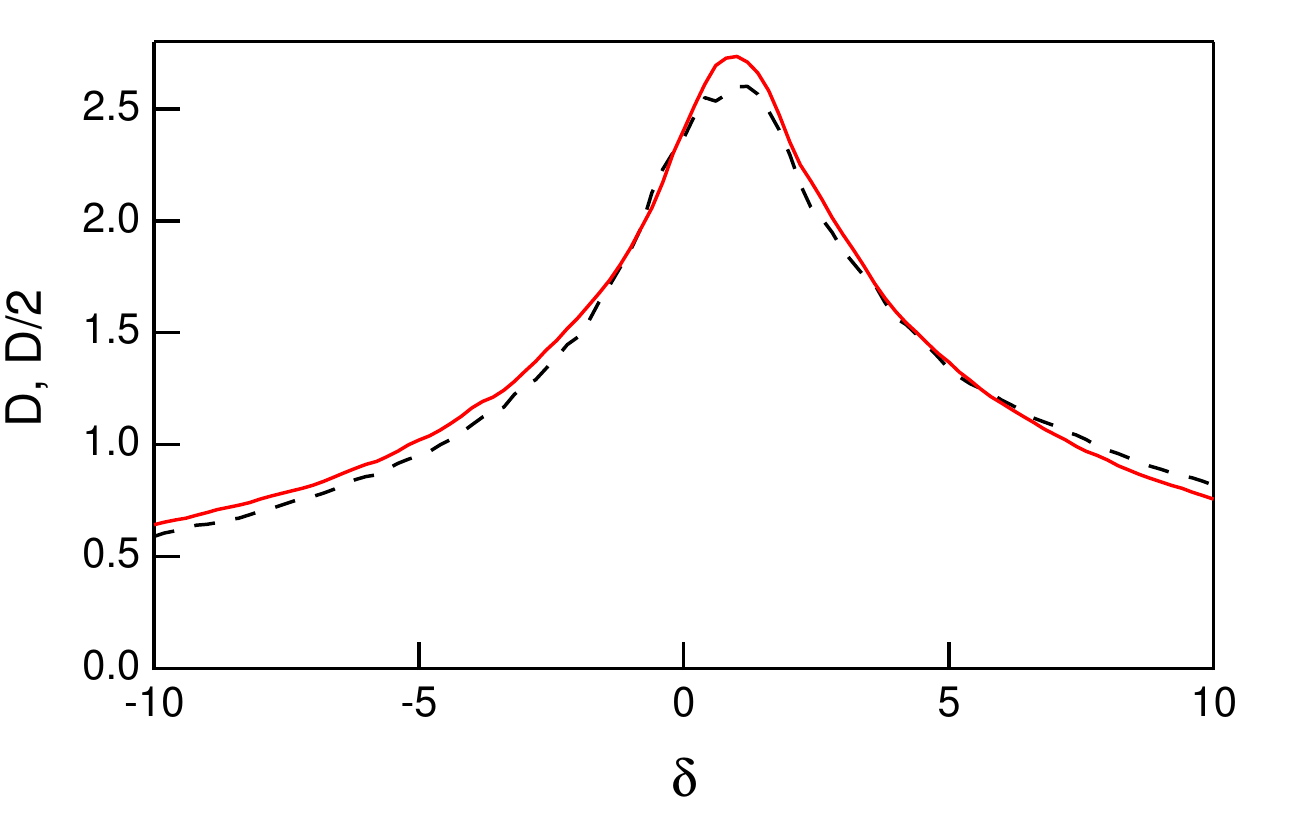}
\caption{Optical thickness as a function of detuning for the sample thickness $h=1$ (solid red line), and half of the optical thickness for a sample twice as thick, $h=2$ (dashed black line). The density $\rho=1$ and the disk area $A=1024$ are the same for both curves.}
\label{THICKSACALING}
\end{figure}

We conclude with an after-the-fact test that reinforces our interpretations. We take simulation data for the same parameters we used to demonstrate the qualitative failure of the MFT in Fig.~\ref{SLABTRANSMISSION}, optical thickness $D$ for the sample density $\rho=1$ and thickness $h=1$ as a function of the tuning of the driving light $\delta$, and plot on the same figure also the optical thickness divided by two for a disk that is twice as thick. The result is shown in Fig.~\ref{THICKSACALING}. The curve with $h=2$ is beset with visible numerical noise since with these atom numbers, $N=2048$, the runs are getting expensive and we have used a reduced number of samples for the atomic positions. Nevertheless, the obvious conclusion is that doubling the thickness from $h=1$ to $h=2$ to a good approximation doubles the optical thickness.

This firstly means that two slabs of thickness $h=1$ back to back would basically behave like one slab with thickness $h=2$. In other words, $h=1$ already represents bulk, 3D, behavior, as we have concluded three times already under different conditions.

Second, within our shortcut to compute the transmission of light,  for these parameters the transmission still decreases exponentially with sample thickness. Now, in such a sample the excitation of the dipoles obviously decrease approximately exponentially downstream in the sample as well. If there were a transition to diffuse optics or if the optics of the finite-size sample ($A=1024$) drastically changed with the increasing thickness, there would be changes in the functional dependence of the dipole moments on the distance downstream, which even our approximate way of calculating the transmission would presumably have picked up. MFT fails when diffuse optics sets in, and the forward-scattering approximation does not fully include the optics of the finite-size disk, but neither of these complications apparently is substantial even for the thicker sample with the maximum optical thickness of $D\simeq5$. We again surmise that our simulations are in the regime when our comparisons with MFT are meaningful.

\section{Concluding remarks}
\label{REMARKS}
Our basic observation is that textbook EDPM  and the ensuing usual optics may fail qualitatively as dipole-dipole interactions between the atoms get stronger with increasing density of the atoms~\cite{JavanainenMFT}.  An effective-medium MFT that spreads out the neighboring atoms into a continuous polarization no longer suffices to describe the influence of the other atoms on an each ``spectator'' atom. Instead, the effect of the other atoms depends on where exactly they are.

In several examples we have studied the question of when the MFT starts showing strain. The scaling of the whole problem we have employed throughout this paper suggests that the relevant scale is on the order $\rho\simeq1$, or $\rho\simeq k^3$ in dimensional units. This is the kind of a density one would see in experiments with Bose-Einstein condensates, or with tightly trapped cold atoms. However, the scaling-away of dimensional quantities in itself does not give any particular numerical criterion. The observation from our simulations is that MFT may be off by quite a lot already at $\rho\sim0.01$.

Here we would like to force the issue of (on-resonance) optical thickness versus density~\cite{JavanainenMFT}, $6\pi\rho h$ versus $\rho$ in our examples.  Optical thickness is a characteristic dimensionless parameter of MFT, and while MFT remains valid, optical thickness may be expected to be {\em the} dimensionless parameter. Numerous theoretical analyses are phrased in terms of optical thickness, and a scaling with optical thickness has been demonstrated in recent experiments, e.g.~\cite{wilkowski2,Guerin_subr16,Ye2016,Araujo16}. It is not a surprise that one can observe superradiance even in standard optics, and that it scales with optical thickness; optical thickness makes optical resonances broader, whereupon the conventional wisdom about Fourier transformations automatically predicts shortertening time scales. However, from our perspective the more interesting case would be when MFT fails, whereupon, we hypothesize, the density becomes an independent parameter governing the deviations from the MFT. We demonstrate such behavior in Fig.~\ref{THICKNESSVARIATION} obtained from our simulations, but at present there apparently are no real experiments showing this type of $\rho$ scaling. On the contrary,  the scaling of subradiance with optical thickness  in a dilute sample as observed experimentally~\cite{Guerin_subr16} severely challenges our picture, as it is unclear if subradiance can exist in MFT in the first place. At the moment we have no resolution to this issue.

There are phenomena for which EDPM and standard optics with their continuous polarization field do not apply as a matter of principle. Incoherent scattering sideways, as in the two bands at the base of the angular distribution of forward scattering on the right panel of Fig.~\ref{ANGULAR}, is an example. In this case, though, we could amend standard optics and still make predictions for sideways scattering: In the single-scattering approximation we would simply add the intensities (not amplitudes) of the light scattered from different atoms. On the other hand, if one studies resonance fluorescence from a few ions, EDPM is a meaningless as a starting point. One can easily imagine intermediate scenarios. What are the predictions from standard optics may also be very difficult to determine per se: If the atomic sample is comparable to the wavelength in size, EDPM boils down to solving the full Maxwell's equations, which remains a challenge even numerically.  All of these caveats notwithstanding, we propose the criterion that a phenomenon should not be called cooperative if standard optics cannot reasonably be excluded as the cause. To give an example, we  would object to the notion that the functioning of eyeglasses reflects cooperative response to light of the molecules that make the lenses.

As we have already noted, our recent interest in this research area was triggered by our observation that we did not see the predicted LL shift in numerical simulations of disks of dense, cold gas. The absence of the LL shift has since been demonstrated in light scattering experiments from a small and dense trapped cloud of atoms~\cite{Pellegrino2014a,Jenkins_thermshift,JEN_longpaper}. These experiments were about sideways scattering, however, which does not directly belong to the MFT framework. From our present viewpoint it is particularly relevant that experiments have also been carried out with forward scattered light under similar conditions that {\em could\/} be directly compared with optics solved numerically from Maxwell's equations~\cite{Jennewein_trans}. The general result was that at higher atom numbers ($\sim 180$) ab-initio simulation analogous to the ones we have described here came closer to the experimental results than the predictions from optics. However, ``[t]he remaining difference with the microscopic model shows that a quantitative understanding of the light-induced interactions even in a relatively simple situation is still a challenge''~\cite{Jennewein_trans}.

The line shifts still present a puzzle. We found the oscillatory dependence of the line shift in accordance with the etalon effects in our simulations of both dilute and inhomogeneously broadened samples, but the LL shift is a more delicate affair. Dimensional analysis  and the experience in spectroscopy suggest that at asymptotically low densities there should be a line shift proportional to sample density $\rho$ ($\propto\rho k^{-3}$ in terms of full dimensional quantities). The LL shifts amounts to a specific prediction for the numerical factor that cannot be deduced from dimensional analysis alone. In dilute homogeneously broadened samples we found a LL type shift that is on the order of $\rho$, but even has the opposite sign than the LL shift. Now, if we expand susceptibility of the gas as a power series in density, the LL shift produces a term proportional to $\rho^2$. In the usual way of MFTs, EDPM apparently is not a systematic expansion in density~\cite{Ruostekoski1997a}. However, there are indications that going beyond MFT in an ensemble of randomly distributed atoms could produce corrections proportional to $\rho^2$ in quantities such as susceptibility~\cite{Morice1995a,Ruostekoski1997a,Ruostekoski1999a} as well, and corresponding density-dependent line shifts. Our tentative conclusion is that, if there is any validity to the usual concept of local-field corrections in homogeneously broadened samples to begin with, beyond-MFT effects probably overwhelm them.

The case of inhomogeneously broadened samples is also intriguing. The low-density phenomenology persisted in our examples at least up to $\rho=1$, which in and of itself is not a surprise as inhomogeneous broadening reduces the dipole-dipole interactions. Qualitatively, only a fraction on the order of the ratio of the homogeneously and inhomogeneously broadened linewidths of the atoms has a frequency that can be on resonance with the light propagating in the sample. This reduces the dipole-dipole interactions and extends the range of validity of the MFT~\cite{Javanainen2014a,Jenkins_thermshift}. On the other hand, when we fitted the Voigt profile to inhomogeneously broadened absorption lines,  we found that the etalon-effect oscillations reside on top of a base lineshift that is about 60\% of the LL shift.

The local-field corrections, of which the LL shift is a particular example, have been an enormously successful concept in the physics of electricity and magnetism for well over a century. The best we could do was to get to within 60\% of the LL shift. If we posit that the notion of local-field corrections is quantitatively sound, the question is, why did we never do better than 60\%? We think that there is a significant piece of physics missing here, but so far it has eluded us.

\acknowledgments
We acknowledge support from NSF, Grant Nos. PHY-0967644 and PHY-1401151, EPSRC, and 2016 Hongik University Research Fund. Most of the computations were done on Open Science Grid, VO Gluex, and on the University of Southampton Iridis 4 computer cluster.

\end{document}